\documentclass[zpreprint,zbstdefault]{zeus_paper}
\usepackage[english]{babel}
\newcommand{\Zdetdesc}{%
A detailed description of the ZEUS detector can be found 
elsewhere~\cite{zeus:1993:bluebook}. A brief outline of the 
components that are most relevant for this analysis is given
below.\xspace}
\xspace
\chardef\usc=95
\chardef\til=126
\catcode`\@=11 % @ signs are now treated as letters
\DeclareRobustCommand\xdotspace{\futurelet\@let@token\@xdotspace}
\def\@xdotspace{%
  \ifx\@let@token.\else
  \ifx\@let@token\bgroup.\else
  \ifx\@let@token\egroup.\else
  \ifx\@let@token\/.\else
  \ifx\@let@token\ .\else
  \ifx\@let@token~.\else
  \ifx\@let@token!.\else
  \ifx\@let@token,.\else
  \ifx\@let@token:.\else
  \ifx\@let@token;.\else
  \ifx\@let@token?.\else
  \ifx\@let@token/.\else
  \ifx\@let@token'.\else
  \ifx\@let@token).\else
  \ifx\@let@token-.\else
  \ifx\@let@token\@xobeysp.\else
  \ifx\@let@token\space.\else
  \ifx\@let@token\@sptoken.\else
   .\space
   \fi\fi\fi\fi\fi\fi\fi\fi\fi\fi\fi\fi\fi\fi\fi\fi\fi\fi}
\catcode`\@=12 % @ signs are no longer letters

\newcommand{\stru}[2]{%
   \relax\ifmmode\hbox{\vrule height#1 depth#2 width0pt}%
   \else\vrule height#1 depth#2 width0pt\fi}
\newcommand{\Ronum}[1]{\uppercase\expandafter{\romannumeral#1}}
\newcommand{\ronum}[1]{\expandafter{\romannumeral#1}}
\DeclareRobustCommand{\LaTeXZ}{%
  \LaTeX\kern-.05em4\kern-.1em
  {\raisebox{-0.2ex}{$\scriptstyle\text{ZEUS}$}}\xspace}
\DeclareMathAlphabet{\mathbf}{OT1}{cmr}{bx}{sl}
\newcommand{\eVdist}{\kern-0.06667em}

\newcommand{\pb}{\,\text{pb}}

\newcommand{\Tesla}{\,\text{T}}

\newcommand{\slashfrac}[2]{%
  \raisebox{0.5ex}{\ensuremath #1}\kern-0.12em/\kern-0.08em
  \raisebox{-.8ex}{\ensuremath #2}}
\newcommand{\sqr}[3]{%
    {\vcenter{\hrule height.#3ex\hbox{\vrule width.#2ex height#1ex
     \kern#1ex\vrule width.#3ex}\hrule height.#2ex}}}

\catcode`\@=11 % @ signs are now treated as letters
\newcommand{\parenbar}{\mathpalette\p@renb@r}
\def\p@renb@r#1#2{\vbox{%
  \ifx#1\scriptscriptstyle \dimen@.7em\dimen@ii.2em\else
  \ifx#1\scriptstyle \dimen@.8em\dimen@ii.25em\else
  \dimen@1em\dimen@ii.4em\fi\fi \offinterlineskip
  \ialign{\hfill##\hfill\cr
    \vbox{\hrule width\dimen@ii}\cr
    \noalign{\vskip-.3ex}%
    \hbox to\dimen@{$\mathchar300\hfil\mathchar301$}\cr
    \noalign{\vskip-.3ex}%
    $#1#2$\cr}}}
\catcode`\@=12 % @ signs are no longer letters

\newcommand{\IP}{{\rm I$\kern-0.01667em$P}\xspace}

\mathchardef\qsm=63
\mathchardef\pls=43
\mathchardef\mns=512
\mathchardef\plm=518
\mathchardef\eql=61
\mathchardef\smallleft=300
\mathchardef\smallright=301
\mathchardef\les=316
\mathchardef\gre=318
\mathchardef\leq=532
\mathchardef\grq=533
\catcode`\@=11 % @ signs are now treated as letters
\newcounter{pict@width}
\newcounter{pict@height}
\newlength{\pict@scale}
\newcommand{\psfigadd}[4]{%
\setcounter{pict@width}{1*\ratio{#2+\pict@scale/2}{\pict@scale}}
\setcounter{pict@height}{1*\ratio{#3+\pict@scale/2}{\pict@scale}}
\setlength{\unitlength}{\pict@scale}
\hbox to #2{\hspace{-\fill}\begin{picture}(\thepict@width,\thepict@height)
\put(0,0){\psfig{figure=#1,width=#2,height=#3,clip=}}
\SetScale{0.283466457}
\SetWidth{1.763889}
{#4}
\end{picture}}
}
\newcounter{pict@widthfst}
\newcounter{pict@widthscd}
\newcounter{pict@widthtot}
\newcommand{\psfigaddtwo}[7]{%
\setcounter{pict@widthfst}{1*\ratio{#2+\pict@scale/2}{\pict@scale}}
\setcounter{pict@widthscd}{1*\ratio{#2+#4+\pict@scale/2}{\pict@scale}}
\setcounter{pict@widthtot}{1*\ratio{#2+#4+#6+\pict@scale/2}{\pict@scale}}
\setcounter{pict@height}{1*\ratio{#3+\pict@scale/2}{\pict@scale}}
\setlength{\unitlength}{\pict@scale}
\hbox{\hspace{-\fill}\begin{picture}(\thepict@widthtot,\thepict@height)
\put(0,0){\psfig{figure=#1,width=#2,height=#3,clip=}}
\put(\thepict@widthscd,0){\psfig{figure=#5,width=#6,height=#3,clip=}}
\SetScale{0.283466457}
\SetWidth{1.763889}
{#7}
\end{picture}}
}
\newcommand{\psfigror}[4]{%
\setcounter{pict@width}{1*\ratio{#2+\pict@scale/2}{\pict@scale}}
\setcounter{pict@height}{1*\ratio{#3+\pict@scale/2}{\pict@scale}}
\setlength{\unitlength}{\pict@scale}
\hbox{\begin{picture}(\thepict@width,\thepict@height)
\put(0,\thepict@height){\psfig{figure=#1,width=#3,height=#2,clip=,angle=270}}
\SetScale{0.283466457}
\SetWidth{1.763889}
{#4}
\end{picture}}
}
\newcommand{\psfigrol}[4]{%
\setcounter{pict@width}{1*\ratio{#2+\pict@scale/2}{\pict@scale}}
\setcounter{pict@height}{1*\ratio{#3+\pict@scale/2}{\pict@scale}}
\setlength{\unitlength}{\pict@scale}
\hbox{\begin{picture}(\thepict@width,\thepict@height)
\put(0,0){\psfig{figure=#1,width=#3,height=#2,clip=,angle=90}}
\SetScale{0.283466457}
\SetWidth{1.763889}
{#4}
\end{picture}}
}
\catcode`\@=12 % @ signs are no longer letters
\newlength\listtextwidth

\catcode`\@=11 % @ signs are now treated as letters
\newlength{\@tabfninsert}
\newlength{\@tabfnwidth}
\newcommand{\tabfootnote}[2]{%
  \setlength{\@tabfninsert}{0.8em}
  \setlength{\@tabfnwidth}{\textwidth}
  \addtolength{\@tabfnwidth}{-\@tabfninsert}
  \addtolength{\@tabfnwidth}{-0.4em}
  \noindent\makebox[\@tabfninsert][r]{\footnotesize$^{#1}$\hfil}\hfill%
  \parbox[t]{\@tabfnwidth}{\footnotesize #2\hfill}}
\catcode`\@=12 % @ signs are no longer letters

\def\etaphi{\eta-\varphi}

\def\sq2{d\sigma/d\q2}
 
\def\q2{Q^2}

\def\pb1{pb$^{-1}$}
\def\g2{GeV$^2$}

\def\mj{M_{\rm jj}}

\def\kt{k_T}

\def\etjb{E^{\rm jet}_{T,{\rm B}}}

\def\etalab{\eta^{\rm jet}_{\rm LAB}}
\def\etlab{E^{\rm jet}_{T,{\rm LAB}}}

\def\colab#1{#1 Coll.}

\def\z0{Z^0}
\def\mz{M_Z}

\def\as{\alpha_s}
\def\oalphas2{{\cal O}(\alpha\as^2)}

\def\oass{{\cal O}(\as^2)}

\def\asz{\as(\mz)}

\def\etal{et al.}

\def\citejetfirst{{\cite{%
epj:c65:363,pl:b560:7,pl:b547:164,pl:b649:12,pl:b653:134,pl:b542:193,*epj:c29:497,*epj:c33:477,*epj:c44:183,*pr:d76:072011,*pr:d78:032004,*np:b792:1,epj:c23:615,np:b765:1,epj:c19:289,*pl:b507:70,pl:b691:127%
}}\xspace}

\def\citebreit{{\cite{%
bookfeynam:1972,*zfp:c2:237%
}}\xspace}

\def\citefits{{\cite{%
zeusfits%
}}\xspace}

\def\citethorbenspaper{{\cite{%
np:b765:1%
}}\xspace}

\def\citeCTD{{\cite{%
nim:a279:290,*npps:b32:181,*nim:a338:254%
}}\xspace}

\def\citeTracking{{\cite{%
nim:a580:1257%
}}\xspace}

\def\citeCAL{\cite{%
nim:a309:77,*nim:a309:101,*nim:a321:356,*nim:a336:23%
}}\xspace

\def\citelumi{\cite{%
desy-92-066,*zfp:c63:391,*acpp:b32:2025%
}}\xspace

\def\citenewlumi{\cite{%
nim:a565:572%
}}\xspace

\xspace

\def\citeelectron{\cite{%
nim:a365:508,*nim:a391:360%
}}\xspace

\def\citeDA{\cite{%
proc:hera:1991:23,*proc:hera:1991:43%
}}\xspace

\def\citektalg{{\cite{%
np:b406:187% 
}}\xspace}

\def\citeinclusive{{\cite{%
pr:d48:3160% 
}}\xspace}

\def\citesnowmass{{\cite{%
proc:snowmass:1990:134%
}}\xspace}

\def\citegeant{{\cite{%
tech:cern-dd-ee-84-1%
}}\xspace}

\def\citesimulation{{\cite{%
zeus:1993:bluebook%
}}\xspace}

\def\citeherakles{{\cite{%
cpc:69:155,*spi:www:heracles%
}}\xspace}

\def\citedjango{{\cite{%
cpc:81:381,*spi:www:djangoh11%
}}\xspace}

\def\citecdm{{\cite{%
pl:b165:147,*pl:b175:453,*np:b306:746,*zfp:c43:625%
}}\xspace}

\def\citeariadne{{\cite{%
cpc:71:15,*zp:c65:285%
}}\xspace}

\def\citelepto{{\cite{%
cpc:101:108%
}}\xspace}

\def\citecteqfive{{\cite{%
epj:c12:375%
}}\xspace}

\def\citecteqsixsix{{\cite{%
Nadolsky:2008zw%
}}\xspace}

\def\citelundstring{{\cite{%
prep:97:31%
}}\xspace}

\def\citejetset{{\cite{%
cpc:82:74,*cpc:39:347,*cpc:43:367,*cpc:135:238%
}}\xspace}

\def\citehector{{\cite{%
cpc:94:128%
}}\xspace}

\def\citenlojet{{\cite{%
nlojet30%
}}\xspace}

\def\citealsnew{{\cite{%
ppnp:58:351%
}}\xspace}

\def\citeelscale{{\cite{%
epj:c21:443%
}}\xspace}

\def\citeem{{\cite{%
epj:c11:427%
}}\xspace}

\begin{document}
%------------------------------------------------------------------------------
%       Title sheet
%------------------------------------------------------------------------------
\prepnum{DESY--10--170}

\title{
Inclusive dijet cross sections in neutral current deep
inelastic scattering at HERA
}                                                       
                    
\author{ZEUS Collaboration}
\draftversion{}
\date{October 2010}

\abstract{
Single- and double-differential inclusive dijet cross sections in neutral current deep inelastic
$ep$ scattering have been measured with the ZEUS detector using 
an integrated luminosity of 374~\pb1. 
The measurement was performed at large values of the photon virtuality, 
$\q2$, between 125 and 20\,000~$\rm{GeV}^2$. The jets were reconstructed
with the $\kt$~cluster algorithm in the Breit reference frame
and selected by requiring their transverse energies in the Breit
frame, $\etjb$, to be larger 
than 8~GeV. In addition, the invariant mass of the dijet system, $\mj$, 
was required to be greater than 20~GeV. 
The cross sections are described by the predictions of next-to-leading-order QCD.
}

\makezeustitle

%------------------------------------------------------------------------------
%       Author List
%------------------------------------------------------------------------------

%===================================================================
%
%  MEMBER NAME  AUTH166 (ZEUS)     M  TEX
%
%  JH.: transformed to a format, which is suited as input for
%       CONVERT, which automatically creates author-indices
%
%  Don't remove lines starting with a percent sign %,
%  CONVERT may need them urgently !
%  
%=====================================================================

%\documentstyle[12pt,twoside]{report}  

%\topmargin-1.cm
%\evensidemargin-0.3cm
%\oddsidemargin-0.3cm
%\textwidth 16.cm
%\textheight 680pt
%\parindent0.cm
%\parskip0.3cm plus0.05cm minus0.05cm
\def\3{\ss}
\pagenumbering{Roman}
                                    % this "%"s are for cosmetics only
%\begin{document}
                                                   %
\begin{center}
{                      \Large  The ZEUS Collaboration              }
\end{center}

{\small

%  members:

{\mbox H.~Abramowicz$^{44, af}$, }
{\mbox I.~Abt$^{34}$, }
{\mbox L.~Adamczyk$^{13}$, }
{\mbox M.~Adamus$^{53}$, }
{\mbox R.~Aggarwal$^{7}$, }
{\mbox S.~Antonelli$^{4}$, }
{\mbox P.~Antonioli$^{3}$, }
{\mbox A.~Antonov$^{32}$, }
{\mbox M.~Arneodo$^{49}$, }
{\mbox V.~Aushev$^{26, aa}$
% ...
}
\newpage

%       institutes:

\makebox[3em]{$^{1}$}
\begin{minipage}[t]{14cm}
{\it Argonne National Laboratory, Argonne, Illinois 60439-4815, USA}~$^{A}$

\end{minipage}\\
\makebox[3em]{$^{2}$}
\begin{minipage}[t]{14cm}
{\it Andrews University, Berrien Springs, Michigan 49104-0380, USA}

\end{minipage}\\
\makebox[3em]{$^{3}$}
\begin{minipage}[t]{14cm}
{\it INFN Bologna, Bologna, Italy}~$^{B}$

\end{minipage}\\
\makebox[3em]{$^{4}$}
\begin{minipage}[t]{14cm}
{\it University and INFN Bologna, Bologna, Italy}~$^{B}$

\end{minipage}\\
\makebox[3em]{$^{5}$}
\begin{minipage}[t]{14cm}
{\it Physikalisches Institut der Universit\"at Bonn,
Bonn, Germany}~$^{C}$

\end{minipage}\\
\makebox[3em]{$^{6}$}
\begin{minipage}[t]{14cm}
{\it H.H.~Wills Physics Laboratory, University of Bristol,
Bristol, United Kingdom}~$^{D}$

\end{minipage}\\
\makebox[3em]{$^{7}$}
\begin{minipage}[t]{14cm}
{\it Panjab University, Department of Physics, Chandigarh, India}

\end{minipage}\\
\makebox[3em]{$^{8}$}
\begin{minipage}[t]{14cm}
{\it Calabria University,
Physics Department and INFN, Cosenza, Italy}~$^{B}$

\end{minipage}\\
\makebox[3em]{$^{9}$}
\begin{minipage}[t]{14cm}
{\it Institute for Universe and Elementary Particles, Chonnam National University,\\
Kwangju, South Korea}

\end{minipage}\\
\makebox[3em]{$^{10}$}
\begin{minipage}[t]{14cm}
{\it Jabatan Fizik, Universiti Malaya, 50603 Kuala Lumpur, Malaysia}~$^{E}$

\end{minipage}\\
\makebox[3em]{$^{11}$}
\begin{minipage}[t]{14cm}
{\it Nevis Laboratories, Columbia University, Irvington on Hudson,
New York 10027, USA}~$^{F}$

\end{minipage}\\
\makebox[3em]{$^{12}$}
\begin{minipage}[t]{14cm}
{\it The Henryk Niewodniczanski Institute of Nuclear Physics, Polish Academy of \\
Sciences, Cracow, Poland}~$^{G}$

\end{minipage}\\
\makebox[3em]{$^{13}$}
\begin{minipage}[t]{14cm}
{\it Faculty of Physics and Applied Computer Science, AGH-University of Science and \\
Technology, Cracow, Poland}~$^{H}$

\end{minipage}\\
\makebox[3em]{$^{14}$}
\begin{minipage}[t]{14cm}
{\it Department of Physics, Jagellonian University, Cracow, Poland}

\end{minipage}\\
\makebox[3em]{$^{15}$}
\begin{minipage}[t]{14cm}
{\it Deutsches Elektronen-Synchrotron DESY, Hamburg, Germany}

\end{minipage}\\
\makebox[3em]{$^{16}$}
\begin{minipage}[t]{14cm}
{\it Deutsches Elektronen-Synchrotron DESY, Zeuthen, Germany}

\end{minipage}\\
\makebox[3em]{$^{17}$}
\begin{minipage}[t]{14cm}
{\it INFN Florence, Florence, Italy}~$^{B}$

\end{minipage}\\
\makebox[3em]{$^{18}$}
\begin{minipage}[t]{14cm}
{\it University and INFN Florence, Florence, Italy}~$^{B}$

\end{minipage}\\
\makebox[3em]{$^{19}$}
\begin{minipage}[t]{14cm}
{\it Fakult\"at f\"ur Physik der Universit\"at Freiburg i.Br.,
Freiburg i.Br., Germany}

\end{minipage}\\
\makebox[3em]{$^{20}$}
\begin{minipage}[t]{14cm}
{\it School of Physics and Astronomy, University of Glasgow,
Glasgow, United Kingdom}~$^{D}$

\end{minipage}\\
\makebox[3em]{$^{21}$}
\begin{minipage}[t]{14cm}
{\it Department of Engineering in Management and Finance, Univ. of
the Aegean, Chios, Greece}

\end{minipage}\\
\makebox[3em]{$^{22}$}
\begin{minipage}[t]{14cm}
{\it Hamburg University, Institute of Experimental Physics, Hamburg,
Germany}~$^{I}$

\end{minipage}\\
\makebox[3em]{$^{23}$}
\begin{minipage}[t]{14cm}
{\it Imperial College London, High Energy Nuclear Physics Group,
London, United Kingdom}~$^{D}$

\end{minipage}\\
\makebox[3em]{$^{24}$}
\begin{minipage}[t]{14cm}
{\it Institute of Particle and Nuclear Studies, KEK,
Tsukuba, Japan}~$^{J}$

\end{minipage}\\
\makebox[3em]{$^{25}$}
\begin{minipage}[t]{14cm}
{\it Institute of Physics and Technology of Ministry of Education and
Science of Kazakhstan, Almaty, Kazakhstan}

\end{minipage}\\
\makebox[3em]{$^{26}$}
\begin{minipage}[t]{14cm}
{\it Institute for Nuclear Research, National Academy of Sciences, and
Kiev National University, Kiev, Ukraine}

\end{minipage}\\
\makebox[3em]{$^{27}$}
\begin{minipage}[t]{14cm}
{\it Kyungpook National University, Center for High Energy Physics, Daegu,
South Korea}~$^{K}$

\end{minipage}\\
\makebox[3em]{$^{28}$}
\begin{minipage}[t]{14cm}
{\it Institut de Physique Nucl\'{e}aire, Universit\'{e} Catholique de Louvain, Louvain-la-Neuve,\\
Belgium}~$^{L}$

\end{minipage}\\
\makebox[3em]{$^{29}$}
\begin{minipage}[t]{14cm}
{\it Departamento de F\'{\i}sica Te\'orica, Universidad Aut\'onoma
de Madrid, Madrid, Spain}~$^{M}$

\end{minipage}\\
\makebox[3em]{$^{30}$}
\begin{minipage}[t]{14cm}
{\it Department of Physics, McGill University,
Montr\'eal, Qu\'ebec, Canada H3A 2T8}~$^{N}$

\end{minipage}\\
\makebox[3em]{$^{31}$}
\begin{minipage}[t]{14cm}
{\it Meiji Gakuin University, Faculty of General Education,
Yokohama, Japan}~$^{J}$

\end{minipage}\\
\makebox[3em]{$^{32}$}
\begin{minipage}[t]{14cm}
{\it Moscow Engineering Physics Institute, Moscow, Russia}~$^{O}$

\end{minipage}\\
\makebox[3em]{$^{33}$}
\begin{minipage}[t]{14cm}
{\it Moscow State University, Institute of Nuclear Physics,
Moscow, Russia}~$^{P}$

\end{minipage}\\
\makebox[3em]{$^{34}$}
\begin{minipage}[t]{14cm}
{\it Max-Planck-Institut f\"ur Physik, M\"unchen, Germany}

\end{minipage}\\
\makebox[3em]{$^{35}$}
\begin{minipage}[t]{14cm}
{\it NIKHEF and University of Amsterdam, Amsterdam, Netherlands}~$^{Q}$

\end{minipage}\\
\makebox[3em]{$^{36}$}
\begin{minipage}[t]{14cm}
{\it Physics Department, Ohio State University,
Columbus, Ohio 43210, USA}~$^{A}$

\end{minipage}\\
\makebox[3em]{$^{37}$}
\begin{minipage}[t]{14cm}
{\it Department of Physics, University of Oxford,
Oxford, United Kingdom}~$^{D}$

\end{minipage}\\
\makebox[3em]{$^{38}$}
\begin{minipage}[t]{14cm}
{\it INFN Padova, Padova, Italy}~$^{B}$

\end{minipage}\\
\makebox[3em]{$^{39}$}
\begin{minipage}[t]{14cm}
{\it Dipartimento di Fisica dell' Universit\`a and INFN,
Padova, Italy}~$^{B}$

\end{minipage}\\
\makebox[3em]{$^{40}$}
\begin{minipage}[t]{14cm}
{\it Department of Physics, Pennsylvania State University, University Park,\\
Pennsylvania 16802, USA}~$^{F}$

\end{minipage}\\
\makebox[3em]{$^{41}$}
\begin{minipage}[t]{14cm}
{\it Polytechnic University, Sagamihara, Japan}~$^{J}$

\end{minipage}\\
\makebox[3em]{$^{42}$}
\begin{minipage}[t]{14cm}
{\it Dipartimento di Fisica, Universit\`a 'La Sapienza' and INFN,
Rome, Italy}~$^{B}$

\end{minipage}\\
\makebox[3em]{$^{43}$}
\begin{minipage}[t]{14cm}
{\it Rutherford Appleton Laboratory, Chilton, Didcot, Oxon,
United Kingdom}~$^{D}$

\end{minipage}\\
\makebox[3em]{$^{44}$}
\begin{minipage}[t]{14cm}
{\it Raymond and Beverly Sackler Faculty of Exact Sciences, School of Physics, \\
Tel Aviv University, Tel Aviv, Israel}~$^{R}$

\end{minipage}\\
\makebox[3em]{$^{45}$}
\begin{minipage}[t]{14cm}
{\it Department of Physics, Tokyo Institute of Technology,
Tokyo, Japan}~$^{J}$

\end{minipage}\\
\makebox[3em]{$^{46}$}
\begin{minipage}[t]{14cm}
{\it Department of Physics, University of Tokyo,
Tokyo, Japan}~$^{J}$

\end{minipage}\\
\makebox[3em]{$^{47}$}
\begin{minipage}[t]{14cm}
{\it Tokyo Metropolitan University, Department of Physics,
Tokyo, Japan}~$^{J}$

\end{minipage}\\
\makebox[3em]{$^{48}$}
\begin{minipage}[t]{14cm}
{\it Universit\`a di Torino and INFN, Torino, Italy}~$^{B}$

\end{minipage}\\
\makebox[3em]{$^{49}$}
\begin{minipage}[t]{14cm}
{\it Universit\`a del Piemonte Orientale, Novara, and INFN, Torino,
Italy}~$^{B}$

\end{minipage}\\
\makebox[3em]{$^{50}$}
\begin{minipage}[t]{14cm}
{\it Department of Physics, University of Toronto, Toronto, Ontario,
Canada M5S 1A7}~$^{N}$

\end{minipage}\\
\makebox[3em]{$^{51}$}
\begin{minipage}[t]{14cm}
{\it Physics and Astronomy Department, University College London,
London, United Kingdom}~$^{D}$

\end{minipage}\\
\makebox[3em]{$^{52}$}
\begin{minipage}[t]{14cm}
{\it Warsaw University, Institute of Experimental Physics,
Warsaw, Poland}

\end{minipage}\\
\makebox[3em]{$^{53}$}
\begin{minipage}[t]{14cm}
{\it Institute for Nuclear Studies, Warsaw, Poland}

\end{minipage}\\
\makebox[3em]{$^{54}$}
\begin{minipage}[t]{14cm}
{\it Department of Particle Physics, Weizmann Institute, Rehovot,
Israel}~$^{S}$

\end{minipage}\\
\makebox[3em]{$^{55}$}
\begin{minipage}[t]{14cm}
{\it Department of Physics, University of Wisconsin, Madison,
Wisconsin 53706, USA}~$^{A}$

\end{minipage}\\
\makebox[3em]{$^{56}$}
\begin{minipage}[t]{14cm}
{\it Department of Physics, York University, Ontario, Canada M3J
1P3}~$^{N}$

\end{minipage}\\
\vspace{30em} \pagebreak[4]

%  references concerning institutes;

\makebox[3ex]{$^{ A}$}
\begin{minipage}[t]{14cm}
 supported by the US Department of Energy\
\end{minipage}\\
\makebox[3ex]{$^{ B}$}
\begin{minipage}[t]{14cm}
 supported by the Italian National Institute for Nuclear Physics (INFN) \
\end{minipage}\\
\makebox[3ex]{$^{ C}$}
\begin{minipage}[t]{14cm}
 supported by the German Federal Ministry for Education and Research (BMBF), under
 contract No. 05 H09PDF\
\end{minipage}\\
\makebox[3ex]{$^{ D}$}
\begin{minipage}[t]{14cm}
 supported by the Science and Technology Facilities Council, UK\
\end{minipage}\\
\makebox[3ex]{$^{ E}$}
\begin{minipage}[t]{14cm}
 supported by an FRGS grant from the Malaysian government\
\end{minipage}\\
\makebox[3ex]{$^{ F}$}
\begin{minipage}[t]{14cm}
 supported by the US National Science Foundation. Any opinion,
 findings and conclusions or recommendations expressed in this material
 are those of the authors and do not necessarily reflect the views of the
 National Science Foundation.\
\end{minipage}\\
\makebox[3ex]{$^{ G}$}
\begin{minipage}[t]{14cm}
 supported by the Polish Ministry of Science and Higher Education as a scientific project No.
 DPN/N188/DESY/2009\
\end{minipage}\\
\makebox[3ex]{$^{ H}$}
\begin{minipage}[t]{14cm}
 supported by the Polish Ministry of Science and Higher Education
 as a scientific project (2009-2010)\
\end{minipage}\\
\makebox[3ex]{$^{ I}$}
\begin{minipage}[t]{14cm}
 supported by the German Federal Ministry for Education and Research (BMBF), under
 contract No. 05h09GUF, and the SFB 676 of the Deutsche Forschungsgemeinschaft (DFG) \
\end{minipage}\\
\makebox[3ex]{$^{ J}$}
\begin{minipage}[t]{14cm}
 supported by the Japanese Ministry of Education, Culture, Sports, Science and Technology
 (MEXT) and its grants for Scientific Research\
\end{minipage}\\
\makebox[3ex]{$^{ K}$}
\begin{minipage}[t]{14cm}
 supported by the Korean Ministry of Education and Korea Science and Engineering
 Foundation\
\end{minipage}\\
\makebox[3ex]{$^{ L}$}
\begin{minipage}[t]{14cm}
 supported by FNRS and its associated funds (IISN and FRIA) and by an Inter-University
 Attraction Poles Programme subsidised by the Belgian Federal Science Policy Office\
\end{minipage}\\
\makebox[3ex]{$^{ M}$}
\begin{minipage}[t]{14cm}
 supported by the Spanish Ministry of Education and Science through funds provided by
 CICYT\
\end{minipage}\\
\makebox[3ex]{$^{ N}$}
\begin{minipage}[t]{14cm}
 supported by the Natural Sciences and Engineering Research Council of Canada (NSERC) \
\end{minipage}\\
\makebox[3ex]{$^{ O}$}
\begin{minipage}[t]{14cm}
 partially supported by the German Federal Ministry for Education and Research (BMBF)\
\end{minipage}\\
\makebox[3ex]{$^{ P}$}
\begin{minipage}[t]{14cm}
 supported by RF Presidential grant N 41-42.2010.2 for the Leading
 Scientific Schools and by the Russian Ministry of Education and Science through its
 grant for Scientific Research on High Energy Physics\
\end{minipage}\\
\makebox[3ex]{$^{ Q}$}
\begin{minipage}[t]{14cm}
 supported by the Netherlands Foundation for Research on Matter (FOM)\
\end{minipage}\\
\makebox[3ex]{$^{ R}$}
\begin{minipage}[t]{14cm}
 supported by the Israel Science Foundation\
\end{minipage}\\
\makebox[3ex]{$^{ S}$}
\begin{minipage}[t]{14cm}
 supported in part by the MINERVA Gesellschaft f\"ur Forschung GmbH, the Israel Science
 Foundation (grant No. 293/02-11.2) and the US-Israel Binational Science Foundation \
\end{minipage}\\
\vspace{30em} \pagebreak[4]

%  references concerning mebers;

\makebox[3ex]{$^{ a}$}
\begin{minipage}[t]{14cm}
also affiliated with University College London,
 United Kingdom\
\end{minipage}\\
\makebox[3ex]{$^{ b}$}
\begin{minipage}[t]{14cm}
now at University of Salerno, Italy\
\end{minipage}\\
\makebox[3ex]{$^{ c}$}
\begin{minipage}[t]{14cm}
now at Queen Mary University of London, United Kingdom\
\end{minipage}\\
\makebox[3ex]{$^{ d}$}
\begin{minipage}[t]{14cm}
also working at Max Planck Institute, Munich, Germany\
\end{minipage}\\
\makebox[3ex]{$^{ e}$}
\begin{minipage}[t]{14cm}
also Senior Alexander von Humboldt Research Fellow at Hamburg University,
 Institute of Experimental Physics, Hamburg, Germany\
\end{minipage}\\
\makebox[3ex]{$^{ f}$}
\begin{minipage}[t]{14cm}
also at Cracow University of Technology, Faculty of Physics,
 Mathemathics and Applied Computer Science, Poland\
\end{minipage}\\
\makebox[3ex]{$^{ g}$}
\begin{minipage}[t]{14cm}
supported by the research grant No. 1 P03B 04529 (2005-2008)\
\end{minipage}\\
\makebox[3ex]{$^{ h}$}
\begin{minipage}[t]{14cm}
now at Rockefeller University, New York, NY
 10065, USA\
\end{minipage}\\
\makebox[3ex]{$^{ i}$}
\begin{minipage}[t]{14cm}
now at DESY group FS-CFEL-1\
\end{minipage}\\
\makebox[3ex]{$^{ j}$}
\begin{minipage}[t]{14cm}
now at Institute of High Energy Physics, Beijing,
 China\
\end{minipage}\\
\makebox[3ex]{$^{ k}$}
\begin{minipage}[t]{14cm}
now at DESY group FEB, Hamburg, Germany\
\end{minipage}\\
\makebox[3ex]{$^{ l}$}
\begin{minipage}[t]{14cm}
also at Moscow State University, Russia\
\end{minipage}\\
\makebox[3ex]{$^{ m}$}
\begin{minipage}[t]{14cm}
now at University of Liverpool, United Kingdom\
\end{minipage}\\
\makebox[3ex]{$^{ n}$}
\begin{minipage}[t]{14cm}
now at CERN, Geneva, Switzerland\
\end{minipage}\\
\makebox[3ex]{$^{ o}$}
\begin{minipage}[t]{14cm}
also affiliated with Universtiy College London, UK\
\end{minipage}\\
\makebox[3ex]{$^{ p}$}
\begin{minipage}[t]{14cm}
now at Goldman Sachs, London, UK\
\end{minipage}\\
\makebox[3ex]{$^{ q}$}
\begin{minipage}[t]{14cm}
also at Institute of Theoretical and Experimental Physics, Moscow, Russia\
\end{minipage}\\
\makebox[3ex]{$^{ r}$}
\begin{minipage}[t]{14cm}
also at INP, Cracow, Poland\
\end{minipage}\\
\makebox[3ex]{$^{ s}$}
\begin{minipage}[t]{14cm}
also at FPACS, AGH-UST, Cracow, Poland\
\end{minipage}\\
\makebox[3ex]{$^{ t}$}
\begin{minipage}[t]{14cm}
partially supported by Warsaw University, Poland\
\end{minipage}\\
\makebox[3ex]{$^{ u}$}
\begin{minipage}[t]{14cm}
now at Istituto Nucleare di Fisica Nazionale (INFN), Pisa, Italy\
\end{minipage}\\
\makebox[3ex]{$^{ v}$}
\begin{minipage}[t]{14cm}
now at Haase Energie Technik AG, Neum\"unster, Germany\
\end{minipage}\\
\makebox[3ex]{$^{ w}$}
\begin{minipage}[t]{14cm}
now at Department of Physics, University of Bonn, Germany\
\end{minipage}\\
\makebox[3ex]{$^{ x}$}
\begin{minipage}[t]{14cm}
also affiliated with DESY, Germany\
\end{minipage}\\
\makebox[3ex]{$^{ y}$}
\begin{minipage}[t]{14cm}
also at University of Tokyo, Japan\
\end{minipage}\\
\makebox[3ex]{$^{ z}$}
\begin{minipage}[t]{14cm}
now at Kobe University, Japan\
\end{minipage}\\
\makebox[3ex]{$^{\dagger}$}
\begin{minipage}[t]{14cm}
 deceased \
\end{minipage}\\
\makebox[3ex]{$^{aa}$}
\begin{minipage}[t]{14cm}
supported by DESY, Germany\
\end{minipage}\\
\makebox[3ex]{$^{ab}$}
\begin{minipage}[t]{14cm}
supported by the Bogolyubov Institute for Theoretical Physics of the National
 Academy of Sciences, Ukraine\
\end{minipage}\\
\makebox[3ex]{$^{ac}$}
\begin{minipage}[t]{14cm}
STFC Advanced Fellow\
\end{minipage}\\
\makebox[3ex]{$^{ad}$}
\begin{minipage}[t]{14cm}
nee Korcsak-Gorzo\
\end{minipage}\\
\makebox[3ex]{$^{ae}$}
\begin{minipage}[t]{14cm}
This material was based on work supported by the
 National Science Foundation, while working at the Foundation.\
\end{minipage}\\
\makebox[3ex]{$^{af}$}
\begin{minipage}[t]{14cm}
also at Max Planck Institute, Munich, Germany, Alexander von Humboldt
 Research Award\
\end{minipage}\\
\makebox[3ex]{$^{ag}$}
\begin{minipage}[t]{14cm}
now at Tokyo Metropolitan University, Japan\
\end{minipage}\\
\makebox[3ex]{$^{ah}$}
\begin{minipage}[t]{14cm}
now at Nihon Institute of Medical Science, Japan\
\end{minipage}\\
\makebox[3ex]{$^{ai}$}
\begin{minipage}[t]{14cm}
now at Osaka University, Osaka, Japan\
\end{minipage}\\
\makebox[3ex]{$^{aj}$}
\begin{minipage}[t]{14cm}
also at \L\'{o}d\'{z} University, Poland\
\end{minipage}\\
\makebox[3ex]{$^{ak}$}
\begin{minipage}[t]{14cm}
member of \L\'{o}d\'{z} University, Poland\
\end{minipage}\\
\makebox[3ex]{$^{al}$}
\begin{minipage}[t]{14cm}
now at Lund University, Lund, Sweden\
\end{minipage}\\
\makebox[3ex]{$^{am}$}
\begin{minipage}[t]{14cm}
also at University of Podlasie, Siedlce, Poland\
\end{minipage}\\

}

%\end{document}

%------------------------------------------------------------------------------
%       Text
%------------------------------------------------------------------------------
\pagenumbering{arabic} 
\pagestyle{plain}
% ----------------------------------------------------------------------------
%       Introduction
% ----------------------------------------------------------------------------
\section{Introduction}
\label{sec-int}

Measurements of jet cross sections are a well established tool for QCD
studies and have been performed for many different observables 
at HERA~\citejetfirst. For jet analyses in neutral current (NC) deep inelastic scattering (DIS), the Breit reference frame~\citebreit is preferred, since it provides a maximal separation between the products
of the beam fragmentation and the hard jets. In this frame, the exchanged space-like virtual boson has 3-momentum ${\bf q}=(0,0,-Q)$ and is
collinear with the incoming parton. While retaining hard QCD processes at leading order in the strong coupling constant $\as$, the contribution from the parton-model
process can be suppressed by requiring the production of jets with high
transverse energy in the Breit frame.
Therefore, measurements of jet cross sections in the Breit frame are directly sensitive to
hard QCD processes, allowing tests of perturbative QCD (pQCD), 
of the factorisation ansatz
and of the parton distribution functions (PDFs) of the proton.

At large boson virtualities, $\q2$, the experimental and theoretical systematic uncertainties are small and, thus, use of the large HERA data sample can provide powerful physical constraints. Jet cross-section 
data from the high-$\q2$ region have been included
in the ZEUS PDF fit, thereby
significantly reducing the uncertainty on the gluon density in the medium- to 
high-$x$ region~\citefits.

Measurements of dijet production in DIS at HERA have so far been performed
with either smaller data sets~\cite{np:b765:1,epj:c19:289,*pl:b507:70} or normalised to the inclusive NC DIS cross section~\cite{epj:c65:363}. 
In this paper, measurements of inclusive dijet production
at large values of $\q2$ are presented using an integrated luminosity of 374~\pb1. Here, differential dijet
cross sections as a function of $\q2$, of the mean jet transverse energy of the dijet system
in the Breit frame, $\overline{\etjb}$, of the dijet invariant mass, $\mj$, of the half-difference of the jet
pseudorapidities in the Breit frame, $\eta^{*} =\left|\eta_{\rm{B}}^{\rm{jet1}}-\eta_{\rm{B}}^{\rm{jet2}}\right|/2$, of the fraction 
of the proton momentum taken by the interacting parton, $\xi = x_{\rm{Bj}}\left(1+\left(\mj\right)^2/\q2\right)$, 
and of $x_{\rm{Bj}}$ are presented.
Here, $x_{\rm{Bj}}$ is the Bjorken scaling variable that defines, for the parton-model process, the fraction
of the proton momentum carried by the struck massless parton. Measurements of the dijet cross section as 
a function of $\xi$ and $\overline{\etjb}$ are also shown for different regions of $\q2$.

The measured single- and double-differential cross sections are compared with next-to-leading-order (NLO) QCD calculations.

% ----------------------------------------------------------------------------
%       Experimental set-up
% ----------------------------------------------------------------------------
\section{Experimental set-up}
\label{sec-setup}

\Zdetdesc

Charged particles were tracked in the central tracking detector
(CTD)~\citeCTD, which operated
in a magnetic field of $1.43\Tesla$ provided by a thin superconducting
solenoid. The CTD consisted of $72$~cylindrical drift-chamber
layers, organised in nine superlayers covering the
polar-angle region \mbox{$15^\circ<\theta<164^\circ$}.
For data taken during the years 1998 to 2000, tracks were reconstructed using the CTD only. 
Starting from the year 2004, the CTD and a silicon microvertex
detector (MVD)~\cite{nim:a581:656}, installed between
the beam-pipe and the inner radius of the CTD, were used. 

The high-resolution uranium--scintillator calorimeter
(CAL)~\citeCAL~covered $99.7\%$ of the total solid angle and consisted of three parts:
the forward (FCAL), the barrel (BCAL) and the rear (RCAL) calorimeters. 
Each part was subdivided transversely into towers and longitudinally
into one electromagnetic section (EMC) and either one (in RCAL) or two
(in BCAL and FCAL) hadronic sections (HAC). The smallest subdivision
of the calorimeter is called a cell. Under test-beam conditions, the
CAL single-particle relative energy resolutions were
$\sigma(E)/E=0.18/\sqrt E$ for (anti-)electrons and 
$\sigma(E)/E=0.35/\sqrt E$ for hadrons, with $E$ in GeV.

The luminosity was measured using the Bethe-Heitler reaction 
$ep\rightarrow e\gamma p$ by the luminosity detector which consisted of a 
lead--scintillator~\citelumi~calorimeter and, in the 2004--2007 running period, an independent 
magnetic spectrometer~\citenewlumi. 

The electron\footnote{Here and in the following, the term ``electron'' denotes 
generically both the electron and the positron unless otherwise stated.} beam 
in HERA was naturally transversely polarised
through the Sokolov-Ternov effect~\cite{spd:8:1203,sjnp:b9:238}.
The characteristic build-up time for the HERA accelerator
was approximately 40 minutes.
Starting from the year 2004, spin rotators on either side of the ZEUS detector
changed the transverse polarisation of the beam
into longitudinal polarisation and back again.
The electron beam polarisation was measured
using two independent polarimeters,
the transverse polarimeter (TPOL)~\cite{nim:a329:79} and
the longitudinal polarimeter (LPOL)~\cite{nim:a479:334}.
Both devices exploited the spin-dependent cross section
for Compton scattering of circularly polarised photons off electrons
to measure the beam polarisation.
The luminosity and polarisation measurements were made over time intervals 
that were much shorter than the polarisation build-up time.

% ----------------------------------------------------------------------------
%       Event Selection
% ----------------------------------------------------------------------------
\section{Event selection and reconstruction}
\label{sec-evtsel}

The data used in this analysis 
were collected during the periods 1998--2000 and 2004--2007, when HERA
operated with protons of energy $E_p=920$~GeV and electrons or
positrons 
of energy $E_e=27.5$~GeV, and
correspond to an integrated luminosity of 203~\pb1 for
the electron data and 171~\pb1 for the positron data. The mean luminosity-weighted
average polarisation of the data was $-0.03$.

A three-level trigger system was used to select events
online~\cite{zeus:1993:bluebook,newtrigger,*proc:chep:1992:222}. At the third
level, NC DIS events were accepted on the basis of the identification
of a scattered-electron candidate using localised energy depositions
in the CAL. At the second level, charged-particle tracks were reconstructed online by using the ZEUS
global tracking trigger~\citeTracking, which combined
information from the CTD and MVD. These online tracks were used to
reconstruct the interaction vertex and reject non-$ep$
background. At the first level, only coarse calorimeter and tracking information was available. Events were selected using criteria based on the energy and transverse energy measured 
in the CAL. Starting from the year 2004, additional tracking requirements were introduced to adapt the trigger rates to the higher instantaneous luminosity.

Events were selected offline using criteria 
which were slightly changed 
with respect to those used in the previous ZEUS dijet measurement~\citethorbenspaper. These changes 
reflect the new phase-space definition of the measurements adopted here. The main steps of the selection are briefly listed
below. 

The scattered-electron candidate was identified from the pattern of
energy deposits in the CAL~\citeelectron. The
energy, $E_e^{\prime}$, the polar\footnote{
The ZEUS coordinate system is a right-handed Cartesian system, with the $Z$
axis pointing in the proton-beam direction, referred to as the ``forward
direction'', and the $X$ axis pointing left towards the center of HERA.
The coordinate origin is at the nominal interaction point.\xspace} angle, $\theta_e$, and the azimuthal angle, $\phi_e$, 
of the electron candidate were determined from the CAL measurements. For events in which 
the electron was found inside the CTD acceptance, the angles $\theta_e$ and $\phi_e$ 
were reconstructed from the associated electron track.
The photon virtuality $\q2$ and the Bjorken scaling variable $x_{\rm{Bj}}$ 
were reconstructed using the double angle (DA)
method~\citeDA. The inelasticity variable $y$ was determined from the condition
$y=\q2/x_{\rm{Bj}}s$, where $s$ is the square of the centre-of-mass energy.

\subsection{Inclusive event selection}
\label{sec-incl-eventsel}
The phase space of the measurement was $125 < \q2 < 20\,000$~\g2 and $0.2 < y < 0.6$.

Events were selected if:
\begin{itemize}
\item an electron candidate of energy $E_{e}^{\prime}>10$~GeV was found.
 This requirement ensured a high and well understood electron-finding
 efficiency and suppressed background from photoproduction events in
 which the scattered electron escapes down the rear beampipe;
\item the total energy not associated with the electron candidate within
  a cone of radius 0.7 units in the pseudorapidity-azimuth ($\etaphi$)
  plane around the electron direction was less than $10\%$ of
  the electron energy. This condition removed photoproduction and DIS events in which part
  of a jet was falsely identified as the scattered electron;
\item a track matched to the energy deposit in the CAL was found in events in which an electron was reconstructed within
the acceptance of the tracking detectors. This was done by restricting the distance of closest approach between the track
extrapolated to the CAL surface and the energy cluster position to within $10~\rm{cm}$, and requiring an electron track momentum
greater than $3~\rm{GeV}$;
\item the vertex position along the beam axis was in the range that 
  was given by the nominal vertex position plus or minus three times the width
  of the vertex distribution approximated by a Gaussian. Both the nominal
  vertex position and the width of the distribution varied between the different 
  data-taking periods. 
  Typical values were
  $|Z_{\rm{vtx}}|<30$~cm. This condition helped to select events consistent 
  with $ep$ interactions;
\item $P_{T,{\rm miss}}/\sqrt{E_T}<2.5$~GeV$^{1/2}$, 
  where $P_{T,{\rm miss}}$ is the
  missing transverse momentum as measured with the CAL
  and $E_T$ is the total transverse energy in the CAL.
  This cut removed cosmic-ray events and beam-related background;
\item $38<(E-P_Z)<65$~GeV, where $E$ is the total energy,
  $E=\sum_iE_i$, and $P_Z$ is the $Z$-component of the vector 
  ${\bf P}=\sum_i {\bf{p_i}}$. The sums run over all clusters of energy deposits in the CAL. 
  This cut removed events with large initial-state radiation and
  further reduced the background from photoproduction.   
\end{itemize}
In addition, events were rejected if a 
second electron candidate with azimuthal separation $\Delta\phi >$~3 from the 
first candidate was found, a ratio of transverse momenta of the two candidates between 0.8 and 1.2
was measured, and, in addition, the rest 
of the CAL energy, besides the two electromagnetic energy clusters, was below
3~GeV. This condition removed elastic Compton scattering events ($ep\rightarrow e\gamma p$).

\subsection{Jet selection}
\label{sec-jet-sel}
The $\kt$ cluster algorithm~\citektalg was used in the
longitudinally invariant inclusive mode~\citeinclusive to
reconstruct jets in the hadronic final state 
assuming massless objects. In data, the 
algorithm was applied to the CAL cells after excluding
those associated with the scattered-electron candidate. The jet search
was performed in the $\etaphi$ plane of the Breit frame. The jet
variables were defined according to the Snowmass
convention~\citesnowmass.

After reconstructing the jet variables in the Breit frame, the
massless four-momenta were boosted into the laboratory frame, where
the transverse energy, $\etlab$, and the pseudorapidity, $\etalab$, of
each jet were calculated. Energy corrections~\cite{pl:b547:164,pl:b558:41,pl:b531:9} were then applied to 
the jets in the laboratory frame and propagated into $\etjb$, 
the transverse jet energy in the Breit frame, in order to compensate for
energy losses in the inactive material in front of the CAL. 

The following cuts were applied to select a clean sample of high-$\q2$ DIS jet events: 
\begin{itemize}
\item events were removed from the sample if the distance $\Delta R$ of any of
  the jets to the electron candidate in the $\etaphi$ plane of the
  laboratory frame was smaller than 1~unit,
  $\Delta R=\sqrt{(\etalab-\eta^e)^2+(\varphi_{\rm LAB}^{\rm jet}-\varphi^e)^2}<1$,
  where $\varphi^e$ and $\eta^e$ are the azimuthal angle and pseudorapidity of
  the scattered-electron candidate, respectively.
  This requirement removed some background from photoproduction and
  improved the purity of the sample;
\item events were removed from the sample if a jet was in
  the backward region of the detector ($\etalab<-1$).
  This requirement removed events in which a radiated photon from the
  electron was misidentified as a jet in the Breit frame;
\item $\etlab > 3$~GeV. 
  This cut removed a small number of jets for which the uncertainty on
  the energy correction was large.
\end{itemize}

The dijet sample was then selected requiring the events to fullfil the following conditions (which also define the dijet phase space of the measurement):
\begin{itemize}
\item at least two jets in the pseudorapidity range $-1<\etalab<2.5$ were found;
\item of these jets, the two 
      with the highest transverse energies 
  in the Breit frame were required to have $\etjb>8$~GeV;
\item the invariant dijet mass, $M_{jj}$,  of the two highest-transverse-energy 
jets in the event was required
  to exceed 20~GeV. This requirement was introduced to make the theoretical calculations infrared insensitive. Despite this cut the NLO QCD calculations in the region $0.65<\eta^{*}<2$ still exhibited infrared sensitivity (Section~\ref{sec-nlo}).
\end{itemize}

The final sample consisted of 31440 dijet events.

% ----------------------------------------------------------------------------
%        MC
% ----------------------------------------------------------------------------
\section{Monte Carlo simulations and data corrections}
\label{sec-mc}
Samples of Monte Carlo (MC) events were generated to determine the response of the
detector to jets of hadrons and the correction factors necessary to
obtain the hadron-level jet cross sections.
The hadron level is defined in terms of hadrons with lifetime $\tau\geq 10$~ps.
The generated events were
passed through the {\sc Geant}-based~\citegeant ZEUS
detector- and trigger-simulation
programs~\citesimulation. They were reconstructed and
analysed by the same program chain as the data.

Neutral current DIS events including radiative effects were simulated
using the {\sc Heracles}~4.6.1~\citeherakles
program with the {\sc Djangoh}~1.1~\citedjango
interface to the hadronisation programs. {\sc Heracles} includes
corrections for initial- and final-state radiation, vertex and
propagator terms, and two-boson exchange. The QCD cascade is simulated
using the colour-dipole model
(CDM)~\citecdm
including the LO QCD diagrams as implemented in
{\sc Ariadne}~4.08~\citeariadne and, alternatively,
with the matrix-element plus parton-shower (MEPS) approach of {\sc Lepto}~6.5~\citelepto. The
CTEQ5D~\citecteqfive proton PDFs were used for these
simulations. Fragmentation into hadrons is performed using the Lund
string model~\citelundstring as implemented in
{\sc Jetset}~7.4~\citejetset.

The jet search was performed on the MC events using the energy
measured in the CAL cells in the same way as for the
data.
The same jet algorithm was also
applied to the final-state particles (hadron level) and to the partons
available after the parton shower simulation (parton level). 

\begin{figure}[p]
\vfill
\setlength{\unitlength}{1.0cm}
\centerline{
\begin{picture} (18.0,15.0)
\put (1.0,0.0){\epsfig{figure=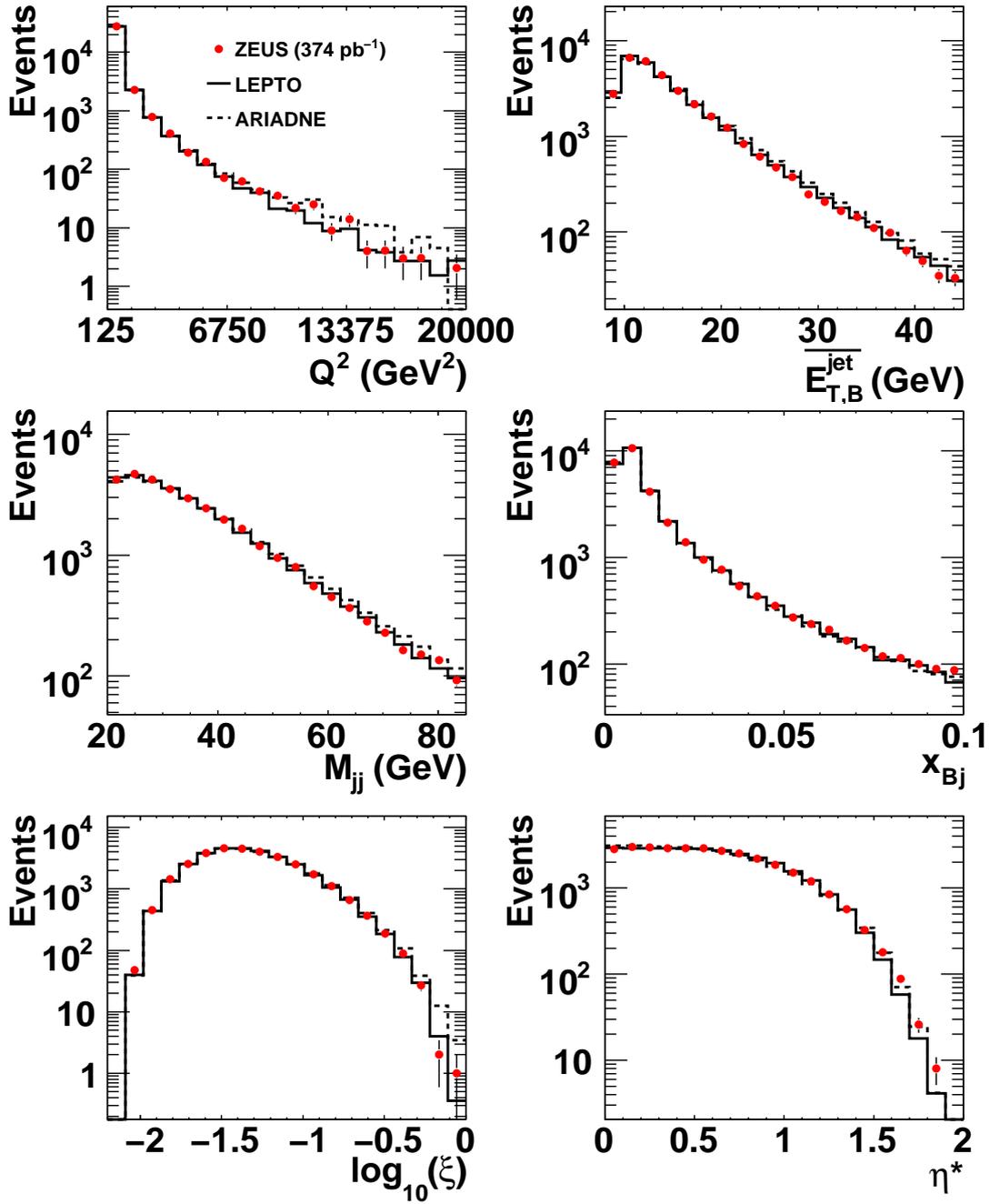,width=0.9\textwidth}}
\end{picture}
}
\caption
{\it Uncorrected data distributions for inclusive dijet production (dots). For comparison, the predictions
of the ARIADNE (dashed histograms) and LEPTO (solid histograms) MC models are also included.}
\label{fig0:0}
\vfill
\end{figure}

\begin{figure}[p]
\vfill
\setlength{\unitlength}{1.0cm}
\centerline{
\begin{picture} (18.0,15.0)
\put (1.0,0.0){\epsfig{figure=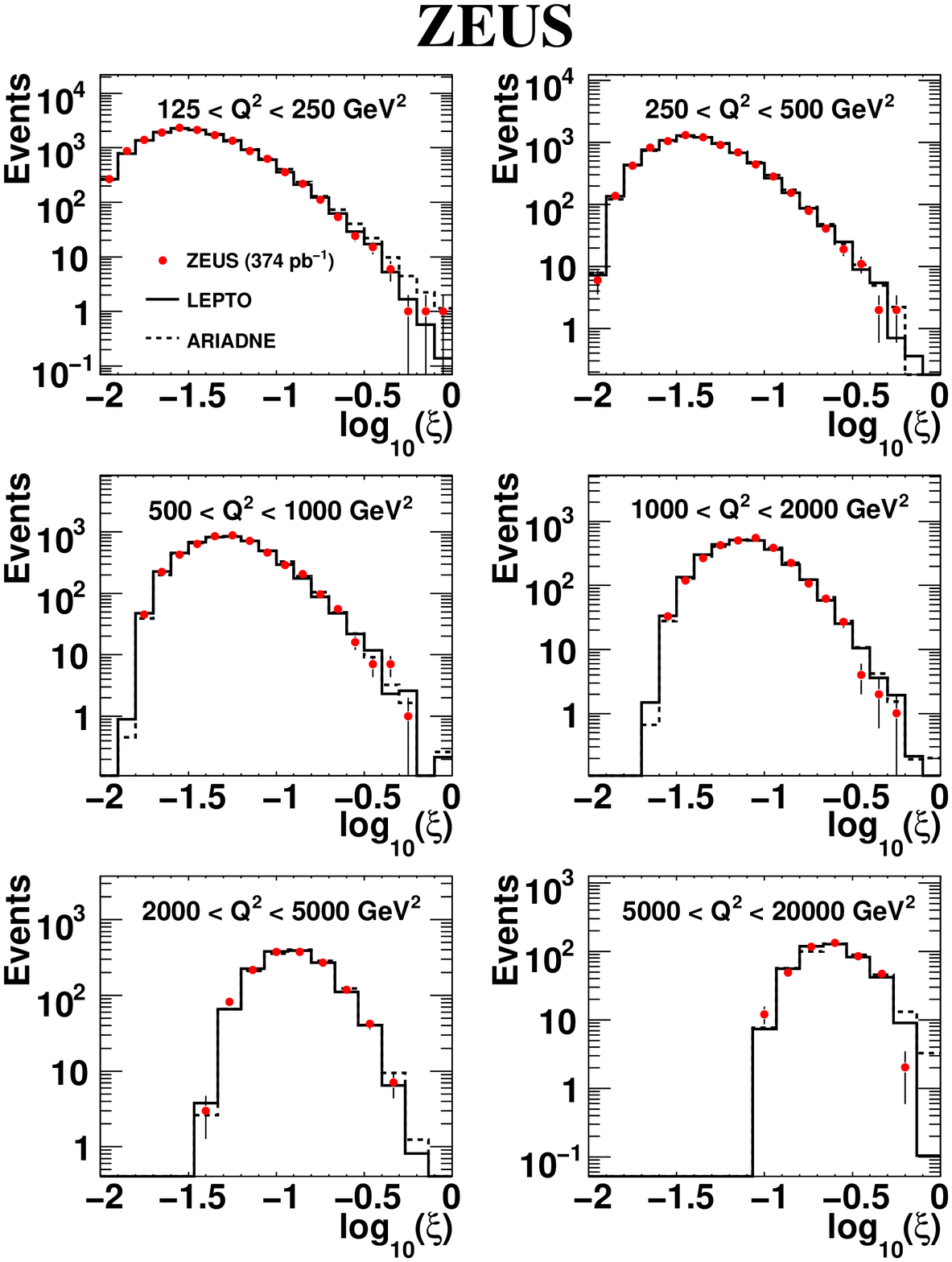,width=0.9\textwidth}}
\end{picture}
}
\caption
{\it Comparison of uncorrected data (dots) and MC model predictions for distributions of $\mathit{\log_{10}\left(\xi\right)}$ in 
  different regions of $\mathit{\q2}$. For comparison, the predictions
of the ARIADNE (dashed histograms) and LEPTO (solid histograms) MC models are also included.}
\label{fig0:1}
\vfill
\end{figure}

\begin{figure}[p]
\vfill
\setlength{\unitlength}{1.0cm}
\centerline{
\begin{picture} (18.0,15.0)
\put (1.0,0.0){\epsfig{figure=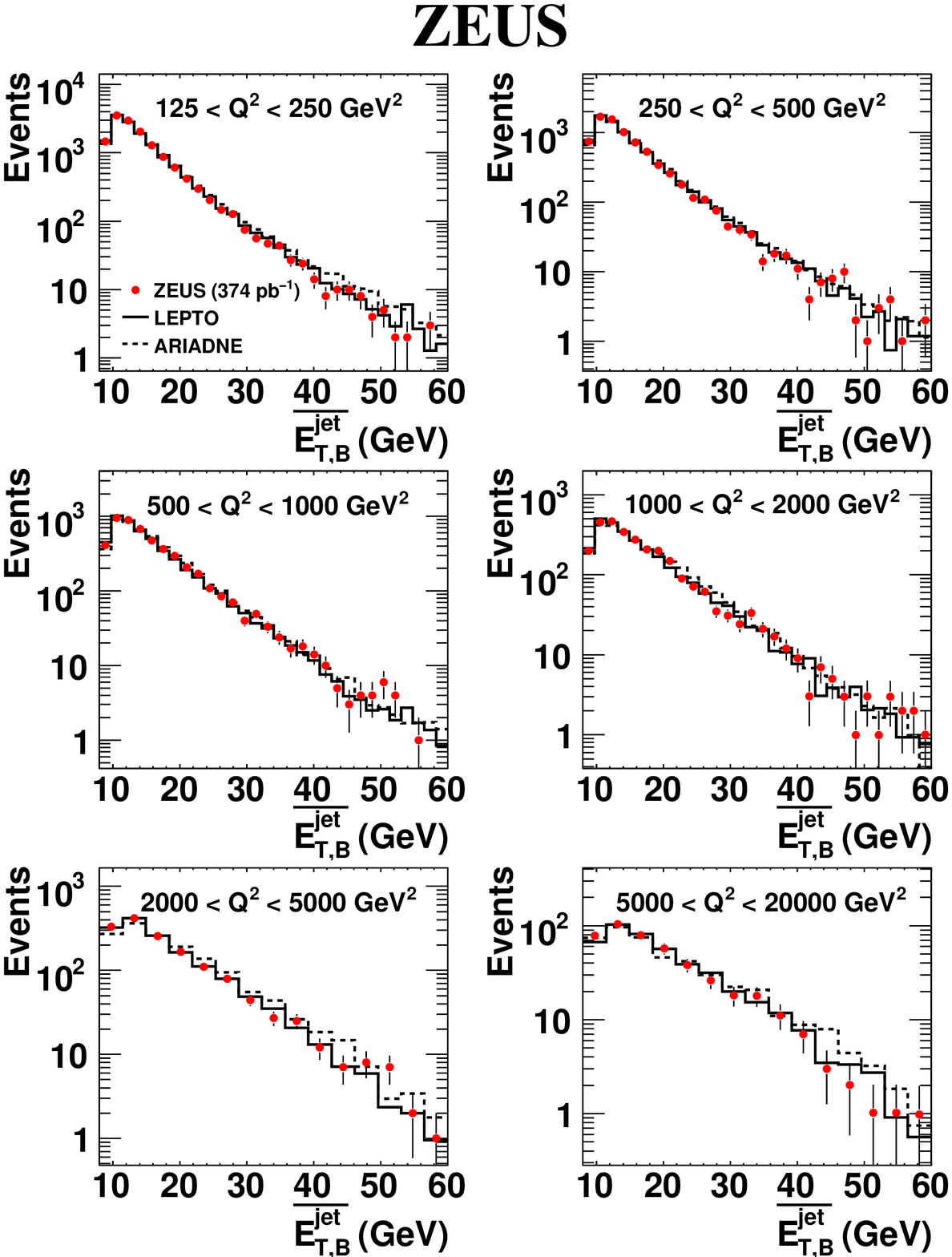,width=0.9\textwidth}}
\end{picture}
}
\caption
{\it Comparison of uncorrected data (dots) and MC model predictions for distributions of $\mathit{\overline{\etjb}}$ in 
different regions of $\mathit{\q2}$. For comparison, the predictions
of the ARIADNE (dashed histograms) and LEPTO (solid histograms) MC models are also included.}
\label{fig0:2}
\vfill
\end{figure}

The data were corrected to the
hadron level and for QED-radiative effects using bin-by-bin correction factors
obtained from the MC samples.
For this approach to be valid, the uncorrected distributions
of the data must be well described by the MC simulations.
This condition was in general satisfied by both the {\sc Ariadne} and
{\sc Lepto} MC. Figures~\ref{fig0:0}-\ref{fig0:2} show 
comparisons of data with MC simulations 
for all observables in which cross sections are presented in this paper.
The MC simulations give a good description of all the data distributions.
The {\sc Lepto} model gives a slightly better
description of the data and was thus used as the default
model; {\sc Ariadne}
was then used to estimate the systematic effect on the
correction procedure due to the parton-shower model.
In all cases, the correction factors differed from unity by
$\left(5-30\right)\%$. These correction factors took into account the efficiency of the
trigger, the selection criteria and the purity and efficiency of the
jet reconstruction. The QED corrections typically amounted to between 3 and 6$\%$. 

Starting in 2004, HERA provided longitudinally polarised lepton beams. The effect of the
polarisation on the measured data events was corrected for by adjusting the data event weights with
the ratio of the predictions for the unpolarised and polarised cross sections as determined
with the {\sc Hector} program~\citehector.

% ----------------------------------------------------------------------------
%        uncertainties
% ----------------------------------------------------------------------------
\section{NLO QCD calculations}
\label{sec-nlo}
Next-to-leading-order ($\oass$) QCD calculations were
obtained using the program {\sc Nlojet++}~\citenlojet. The
calculations were performed in the $\overline{\rm MS}$ renormalisation and
factorisation schemes. The number of flavours
was set to five and the factorisation scale was chosen to be
$\mu_F=Q$. Calculations with different choices of the renormalisation
scale, $\mu_R$, were performed: the default choice
was $\mu_R^2=\q2+\overline{\etjb}^2$. Alternatively, the scales
$\q2$ and $\overline{\etjb}^2$ were investigated.
The strong coupling constant was calculated at two loops with
$\Lambda^{(5)}_{\overline{\rm MS}}=226$~MeV, corresponding to
$\asz=0.118$. The calculations were performed using the
CTEQ6.6~\citecteqsixsix parameterisations of the proton PDFs. The $\kt$ cluster algorithm was also applied to the
partons in the events generated by {\sc Nlojet++} in order to obtain
the jet cross-section predictions.
The predictions of {\sc Nlojet++} were cross-checked with the 
{\sc Disent} program~\cite{np:b485:291}. Both programs agreed to better 
than 1$\%$.

The lack of sensitivity to the infared cutoff of the NLO QCD calculations was verified by determining the total inclusive 
dijet cross section as a function of the $\mj$ cut in each analysis bin separately. Except for the two highest $\eta^{*}$ bins, 
the theoretical predictions were found to be infrared insensitive~\cite{jbehrphdthesis}.

\begin{figure}[p]
\vfill
\setlength{\unitlength}{1.0cm}
\centerline{
\begin{picture} (18.0,15.0)
\put (1.0,0.0){\epsfig{figure=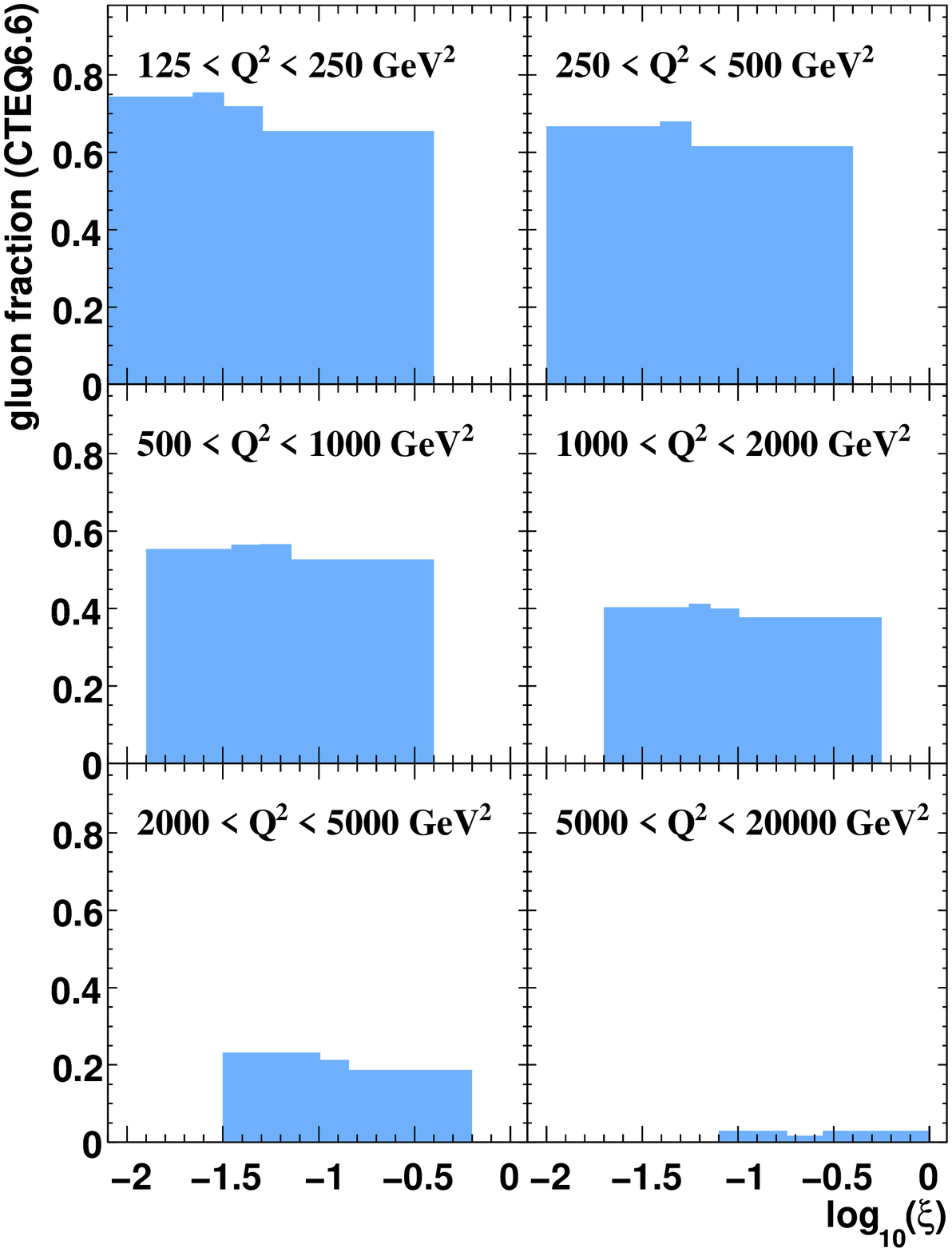,width=0.9\textwidth}}
\end{picture}
}
\caption
{\it The fraction of gluon-induced events as a function of $\mathit{\log_{10}\left(\xi\right)}$ as predicted by the {\sc {\it CTEQ6.6}} PDFs in different regions of $\mathit{\q2}$.}
\label{figgluon_xi}
\vfill
\end{figure}

\begin{figure}[p]
\vfill
\setlength{\unitlength}{1.0cm}
\centerline{
\begin{picture} (18.0,15.0)
\put (1.0,0.0){\epsfig{figure=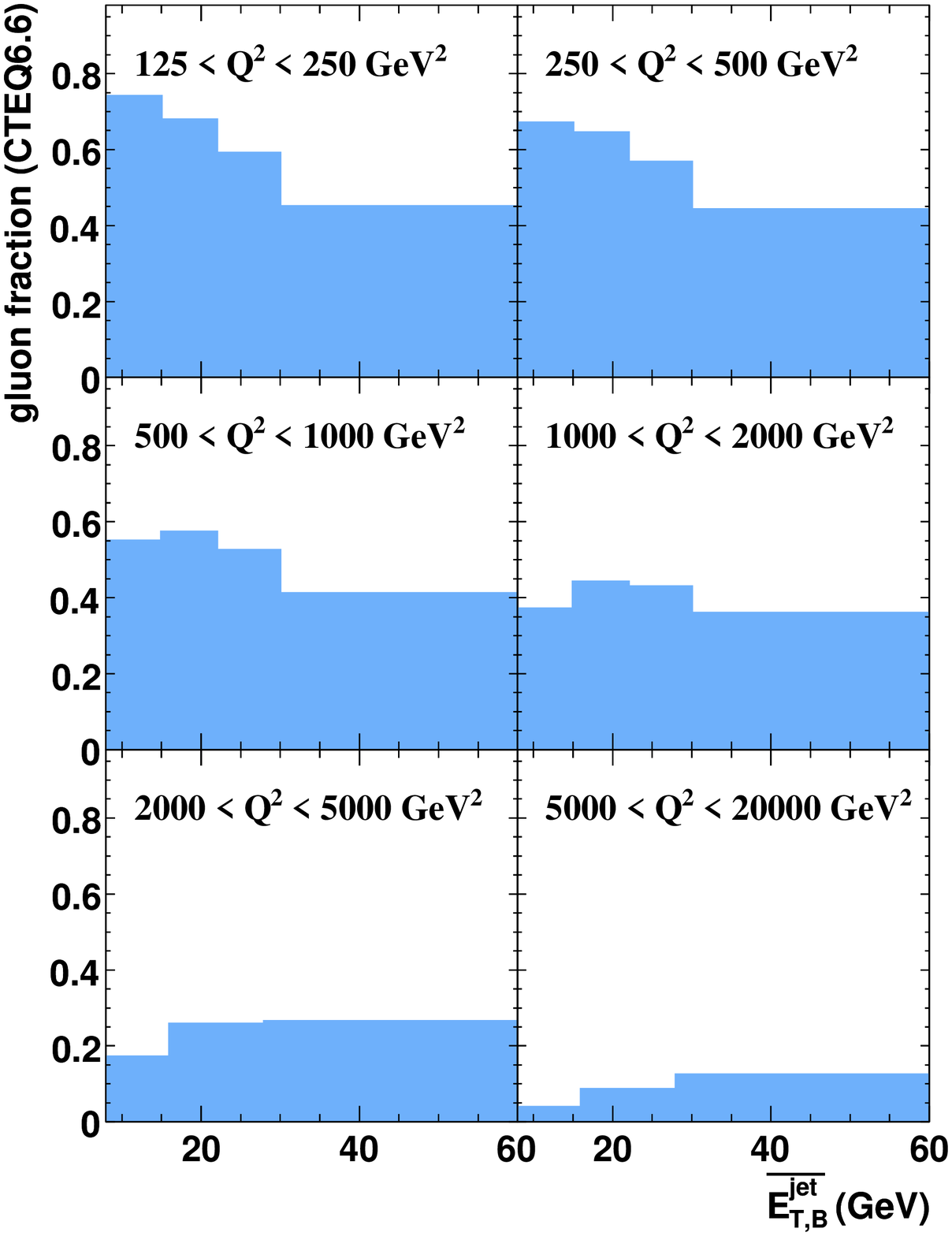,width=0.9\textwidth}}
\end{picture}
}
\caption
{\it The fraction of gluon-induced events as a function of $\mathit{\overline{\etjb}}$ as predicted by the {\sc {\it CTEQ6.6}} PDFs in different regions of $\mathit{\q2}$.}
\label{figgluon_etmean}
\vfill
\end{figure}

\begin{figure}[p]
\vfill
\setlength{\unitlength}{1.0cm}
\centerline{
\begin{picture} (18.0,15.0)
\put (1.0,0.0){\epsfig{figure=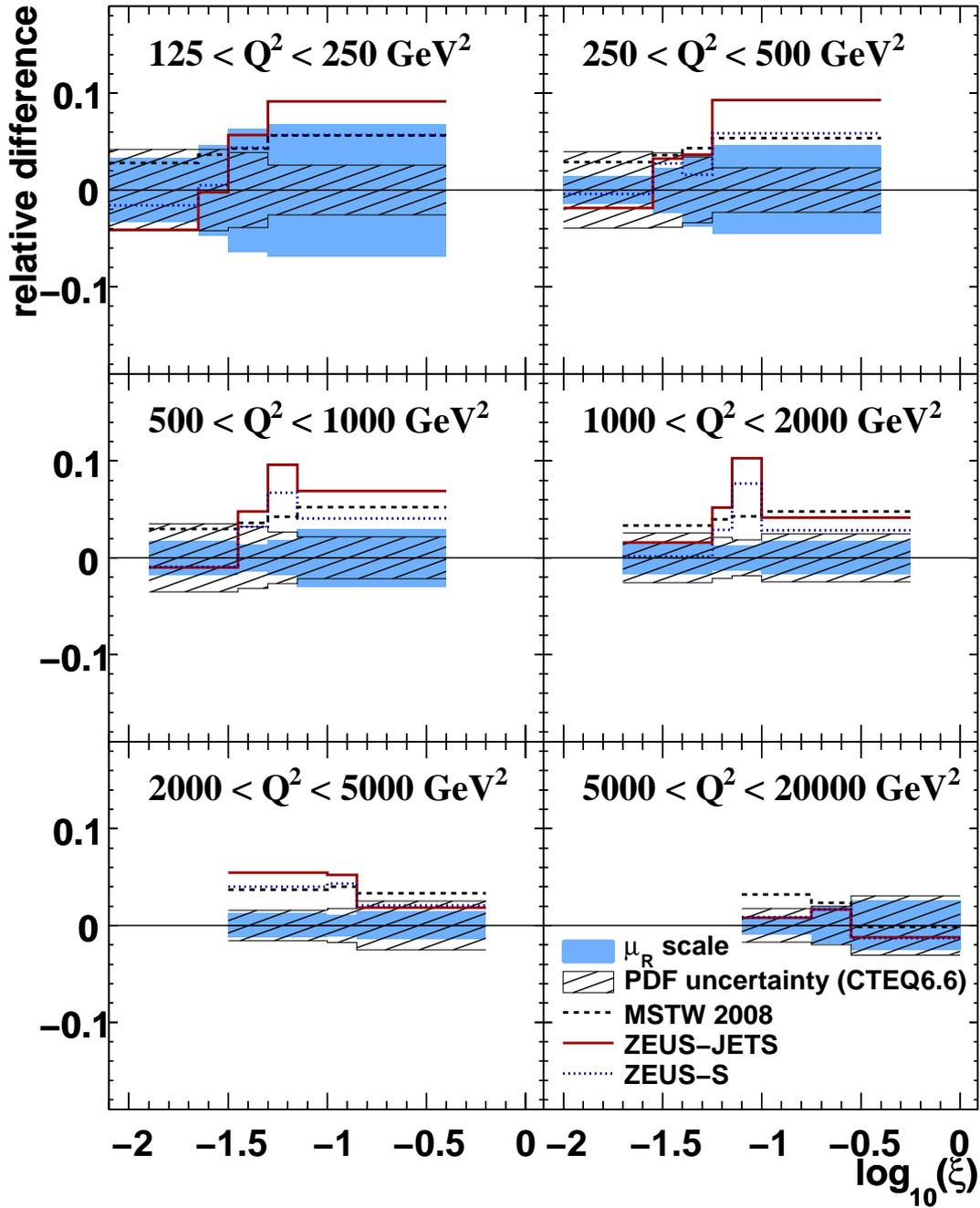,width=0.9\textwidth}}
\end{picture}
}
\caption
{\it The relative {\sc {\it CTEQ6.6}} PDF uncertainty, the relative uncertainty due to missing higher orders estimated by a variation of $\mathit{\mu_{R}}$ and the theoretical predictions from different PDF sets relative to those obtained with {\sc {\it CTEQ6.6}} as functions of $\mathit{\log_{10}\left(\xi\right)}$ in different regions of $\mathit{\q2}$.}
\label{figpdferrors}
\vfill
\end{figure}

The data presented in this paper are, among others, intended for the use in 
QCD PDF fits, aiming specifically at a further improvement of the uncertainty on 
the gluon density at large values of $x$. In order to demonstrate the usefulness 
of the data for this purpose, Figs.~\ref{figgluon_xi} and \ref{figgluon_etmean} show, as a function of the 
variables $\xi$ and $\overline{\etjb}$ in different regions of $\q2$, the NLO predictions for the
fraction of events which are initiated by a gluon from the proton using the {\sc CTEQ6.6} PDFs. This gluon fraction
ranges from about 75$\%$ at 125~$< \q2 <$~250~\g2 and small $\xi$ to
about 5$\%$ at the highest $\q2$~above 5000~\g2, where $\xi$ is approximately 
confined to values above 0.1. In the lower $\q2$~regions, the gluon fraction is also significant for large values of $\xi$. 
Since this region is not statistically limited, precise input for the 
PDF fits can be expected. Figure \ref{figpdferrors} shows the 
relative {\sc CTEQ6.6} PDF uncertainty, the uncertainty due to missing higher orders in the 
calculation estimated by variation of $\mu_{R}$ and the theoretical
predictions from {\sc MSTW2008}~\cite{Martin:2009iq}, {\sc ZEUS-JETS}~\citefits and {\sc ZEUS-S}~\cite{Chekanov:2002pv}. The corresponding uncertainties 
for the latter three PDF sets are not shown. 
The {\sc CTEQ6.6} PDF uncertainty and the observed spread between the various PDF sets is in some regions of the considered 
dijet phase space larger than the uncertainty arising from missing higher orders.

The measurements refer to jets of hadrons, whereas the NLO QCD 
calculations refer to jets of partons. The QCD predictions were 
corrected to the hadron level using the MC hadronisation model 
described in the previous section to give multiplicative factors, $C_{\rm hadr}$, 
defined as the ratio of the cross section for jets of hadrons over that for jets of partons. The ratios obtained
with {\sc Ariadne} and {\sc Lepto} were averaged to obtain the $C_{\rm hadr}$
factors, which differ from unity by less than 5~$\%$.

Neither {\sc Nlojet++} nor {\sc Disent} includes the contributions from $\z0$ exchange;
MC simulated events with and without $\z0$ exchange were used to include this effect in the pQCD predictions.
In the following, pQCD calculations refer to the fully corrected predictions, unless otherwise stated.

Several sources of uncertainty in the theoretical predictions were
considered:
\begin{itemize}
  \item the uncertainty on the NLO QCD calculations due to terms
    beyond NLO, estimated by varying $\mu_R$ by a factor of two up and down,
    was below $\pm$6$\%$ at low $\q2$ and low $\etjb$ and
    decreased to below $\pm$3$\%$ in the highest-$\q2$ region;
  \item the uncertainty on the NLO QCD calculations due to that on
    $\asz$ was estimated by repeating the calculations using two
    additional sets of proton PDFs, CTEQ6.6A3 and
    CTEQ6.6A2, determined assuming $\asz=0.114$
    and $0.122$, respectively. The difference between the calculations
    using these sets and CTEQ6.6 was scaled to
    reflect the current uncertainty on
    $\as$~\citealsnew. 
    The resulting uncertainty on the
    cross sections was mostly below $\pm$3$\%$;
  \item the uncertainty on the modelling of the parton shower was
    estimated as half the difference between the correction
    factors calculated from the {\sc Lepto} and {\sc Ariadne} models.
    The resulting uncertainty on the cross sections was typically less than
    2$\%$;
  \item the uncertainty on the NLO calculations due to the
    proton PDFs was estimated by repeating the
    calculations using 44 additional sets from the CTEQ6.6 analysis, which takes
    into account the
    statistical and correlated systematic experimental uncertainties of each
    data set used in the determination of the proton PDFs.
    The resulting
    uncertainty on the cross sections was about $\pm$4$\%$ at low $\q2$
    and decreased to around $\pm$2$\%$ at high $\q2$;
  \item the uncertainty of the calculations in the value of $\mu_F$ was
    estimated by repeating the calculations with $\mu_F=Q/2$ and $2Q$.
    The effect on the calculations was negligible.
\end{itemize}

The total theoretical uncertainty was obtained by adding in quadrature
the individual uncertainties listed above.

% ----------------------------------------------------------------------------
%        uncertainties
% ----------------------------------------------------------------------------
\section{Experimental uncertainties}
\label{sec-syst}
The following sources of systematic uncertainty were considered for
the measured cross sections \cite{jbehrphdthesis}:
\begin{itemize}
  \item the uncertainty on the absolute energy scale of the jets was
    estimated to be $\pm 1\%$ for $\etlab>10$~GeV 
    and $\pm 3\%$ for lower $\etlab$ values~\cite{epj:c23:615,proc:calor:2002:767,pl:b531:9,jbehrphdthesis}. The 
    resulting uncertainty on the cross sections
    was about $\pm$4$\%$ and increased to
    approximately $\pm$6$\%$ in certain regions of the dijet phase space;
  \item the uncertainty in the absolute energy scale of the electron
    candidate was estimated to be $\pm 1\%$~\citeelscale ($\pm 2\%$~\cite{Chekanov:2009gm}) for the data from the years 1998--2000 (2004--2007).
    The resulting uncertainty was below $\pm 1\%$;
  \item the differences in the results obtained by using either
    {\sc Ariadne} or {\sc Lepto} to correct the data for detector
    effects were typically below $\pm$5$\%$;
  \item the analysis was repeated using an alternative
    technique~\citeem to select the scattered-electron
    candidate. The resulting uncertainty was typically below $\pm$1$\%$;
  \item the $\etlab$ cut was changed to 2 and 4~GeV. The resulting uncertainty was mostly smaller than $\pm$1$\%$;
  \item the uncertainty due to the selection cuts was estimated by
    varying the values of the cuts within the resolution of each
    variable. The effect on the cross sections was in general below
    $\pm 2\%$;
\item  the combined, luminosity-weighted
  systematic error on the polarisation measurement was $3.9\%$. The effect on the cross sections
  was negligible;
\item the simulation of the first-level trigger was corrected in order to match the measured efficiency in the data. The systematic effect 
on the cross sections was typically less than $1\%$.
  \end{itemize}

\begin{figure}[p]
\vfill
\setlength{\unitlength}{1.0cm}
\centerline{
\begin{picture} (18.0,15.0)
\put (1.0,0.0){\epsfig{figure=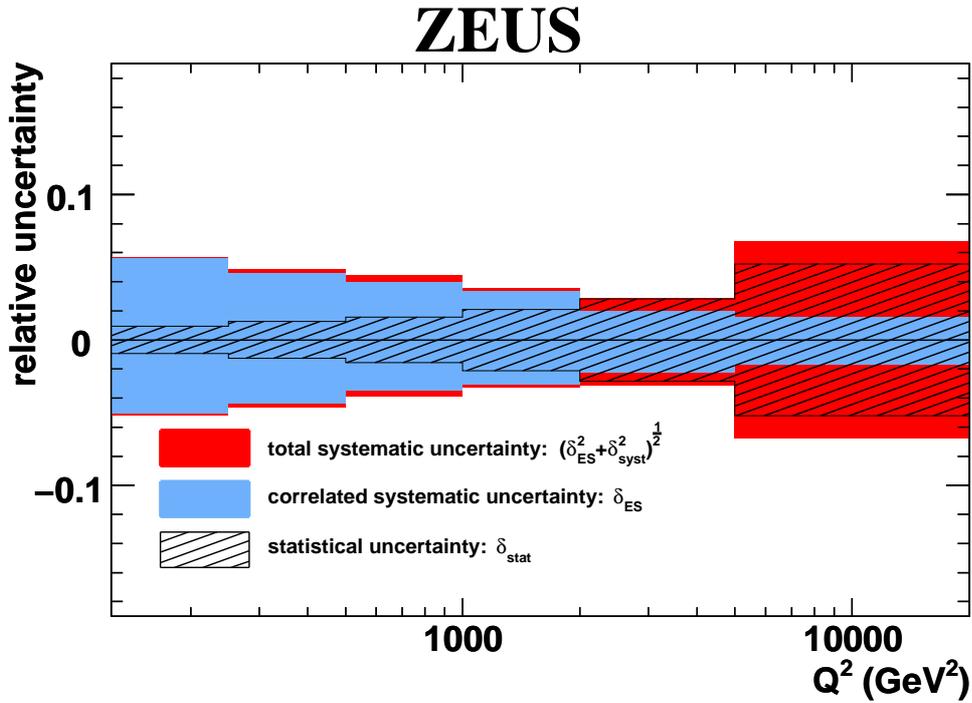,width=0.9\textwidth}}
\end{picture}
}
\caption
{\it The statistical uncertainty, $\mathit{\delta_{\rm{stat}}}$, the correlated uncertainty associated with the energy scale of the jets, $\mathit{\delta_{\rm{ES}}}$, and the quadratic sum of the correlated and uncorrelated, $\mathit{\delta_{\rm{syst}}}$, systematic uncertainties, $\mathit{\sqrt{\delta_{\rm{ES}}^{2}+\delta_{\rm{syst}}^{2}}}$, as functions of $\mathit{\q2}$.}
\label{fig_exp_unc}
\vfill
\end{figure}

The systematic uncertainties not associated with the absolute energy
scale of the jets were added in quadrature. Figure \ref{fig_exp_unc} shows the statistical uncertainty, the correlated systematic uncertainty which is caused by the jet energy scale 
and the quadratic sum of the correlated and uncorrelated systematic uncertainties as a function of $\q2$. Except 
for the high-$Q^{2}$ region, the correlated uncertainty was the
dominating contribution to the total experimental uncertainty.
In addition, there was an
overall normalisation uncertainty of $\pm$2.2$\%$ for the 1998--2000 data and of $\pm$2.6$\%$ for the 2004--2007 data.
Therefore, the combined, luminosity-weighted average systematic uncertainty on the luminosity measurement
was $\pm$2.5$\%$, which was not included in the cross-section figures or the tables.

% ----------------------------------------------------------------------------
%        results
% ----------------------------------------------------------------------------
\section{Results}
\label{sec-results}
The differential inclusive dijet cross sections were measured in the kinematic 
region $125<\q2<20\,000~\rm{GeV}^{2}$ and $0.2 < y < 0.6$. The jets were reconstructed using 
the $\kt$ cluster algorithm in the longitudinally invariant inclusive mode and the cross sections refer to 
jets with $E_{T,\rm{B}}^{\rm{jet}}>8~\rm{GeV}$ and $-1<\etalab<2.5$. The invariant dijet mass of the two highest-transverse-energy jets in the event was required 
to be greater than 20~GeV. These cross sections were corrected for detector and QED radiative effects and the 
running of $\alpha_{\rm{em}}$.

\subsection{Single-differential dijet cross sections}
\label{sec-results-single}
\begin{figure}[p]
\vfill
\setlength{\unitlength}{1.0cm}
\begin{picture} (18.0,15.0)
\put (1.0,0.0){\epsfig{figure=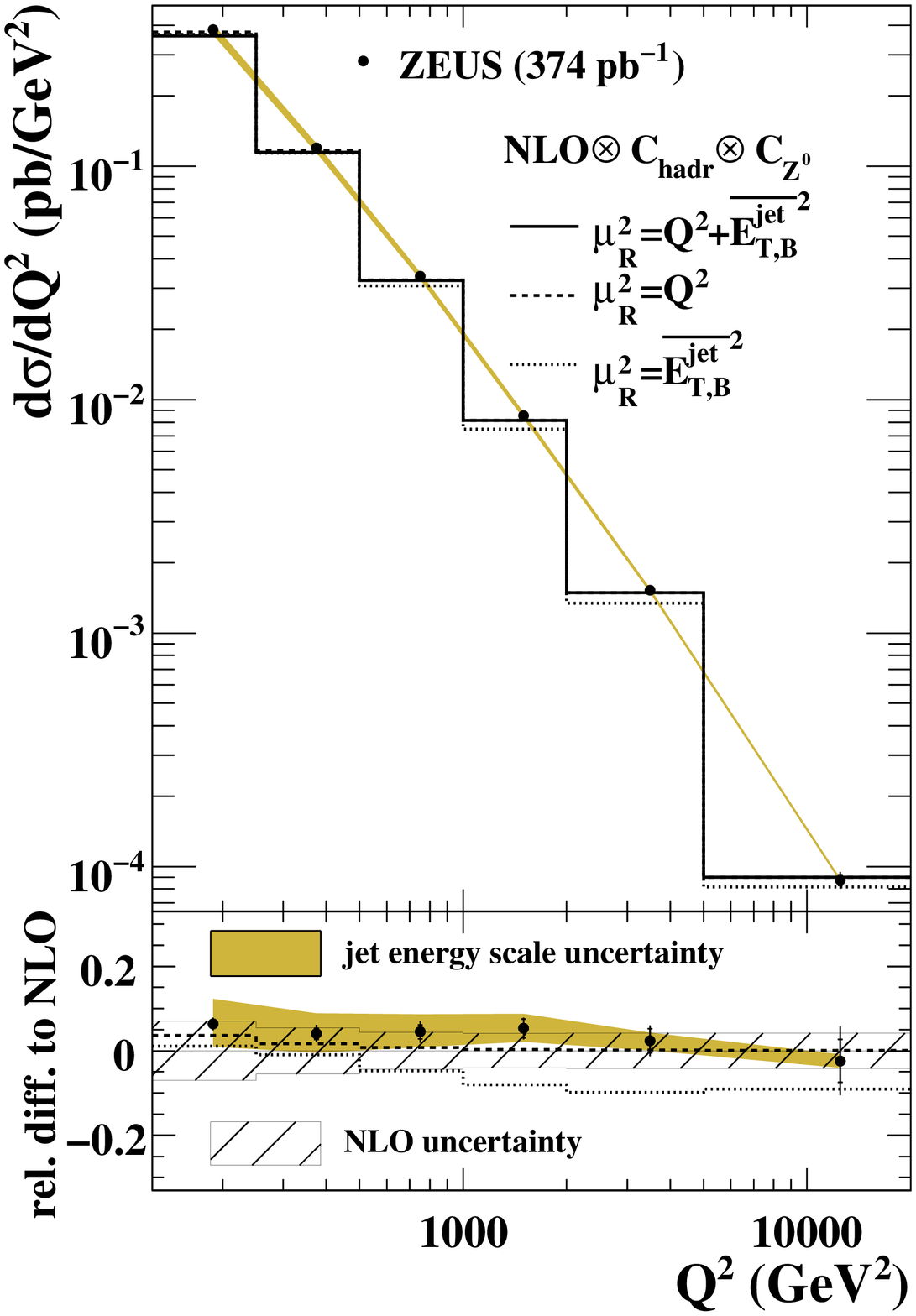,width=7.5cm}}
\put (8.5,0.0){\epsfig{figure=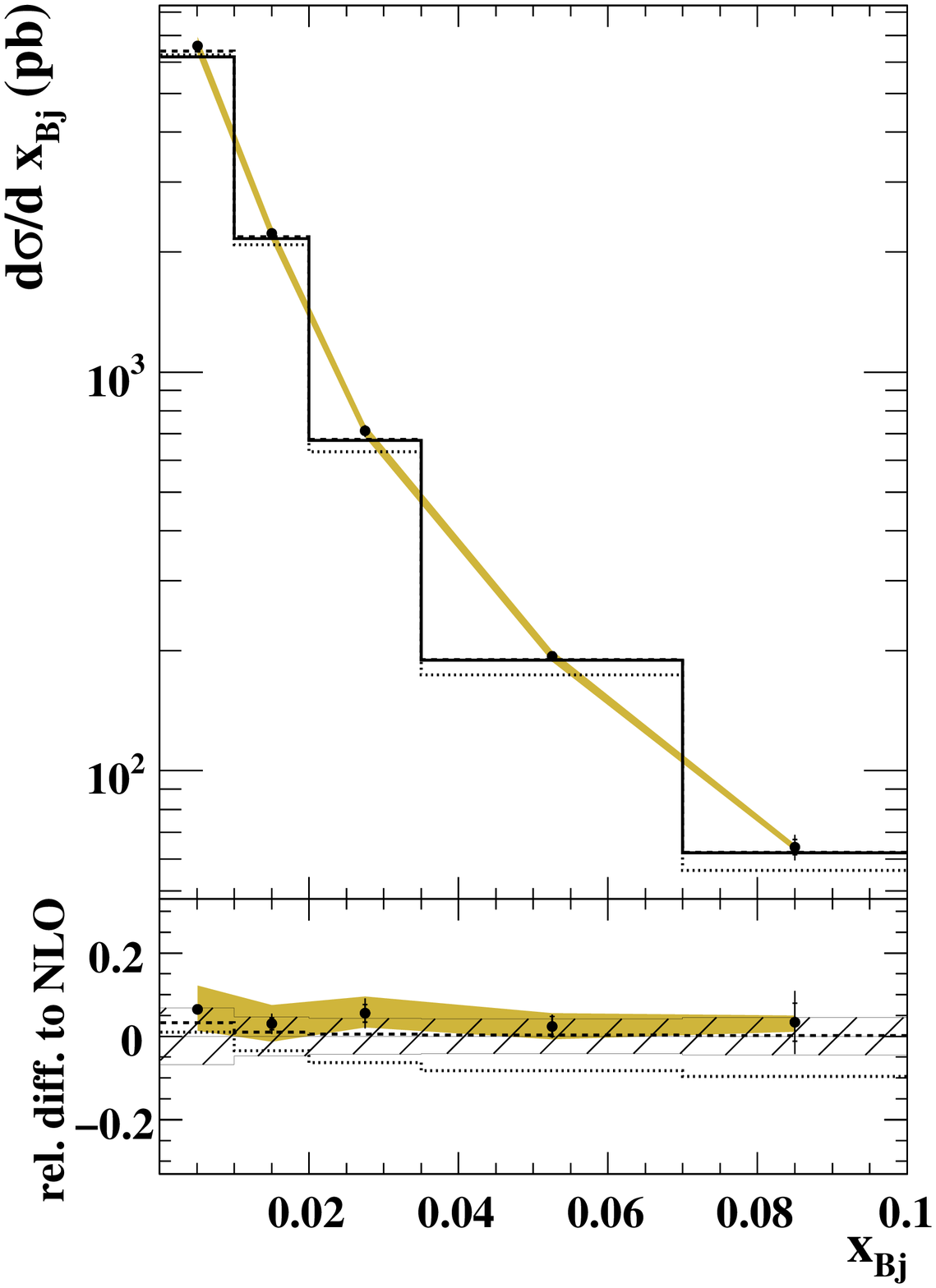,width=7.5cm}}
\put (7.25,10.5){{\bf {\Huge ZEUS}}}
\put (2.8,3.9){{\bf (a)}}
\put (10.3,3.9){{\bf (b)}}
\end{picture}
\caption
{\it The measured differential cross-sections (a) $\mathit{\sq2}$ and (b)
  $\mathit{d\sigma/dx_{{\rm Bj}}}$
for inclusive dijet production with $\mathit{\etjb>8}$~GeV, $\mathit{\mj >}$~20~GeV and $\mathit{-1<\etalab<2.5}$
(dots), in the kinematic range given by 0.2~$\mathit{< y <}$~0.6 and 125~$\mathit{< Q^2
<}$~20\,000~$\rm{\textit{GeV~}}^{\mathit{2}}$. The inner error bars represent the statistical
uncertainty. The outer error bars show the statistical and systematic
uncertainties, not associated with the uncertainty on the absolute energy
scale of the jets, added in quadrature. The shaded bands display the
uncertainties due to the absolute energy scale of the jets. The NLO QCD
calculations with $\mathit{\mu_R^2=\q2+\overline{\etjb}^2}$ (solid lines), $\mathit{\mu_R^2=\q2}$
(dashed lines) and \mbox{$\mathit{\mu_R^2=\overline{\etjb}^2}$} (dotted lines), corrected
for hadronisation effects and $\mathit{\z0}$ exchange and using the {\sc {\it CTEQ6.6}} parameterisations of the
proton PDFs, are also shown. The lower parts of the figures show the
relative differences with respect to the NLO QCD calculations with $\mathit{\mu_R^2=\q2+\overline{\etjb}^2}$. 
The hatched bands display the total theoretical uncertainty. 
}
\label{fig1}
\vfill
\end{figure}

\begin{figure}[p]
\vfill
\setlength{\unitlength}{1.0cm}
\begin{picture} (18.0,15.0)
\put (1.0,0.0){\epsfig{figure=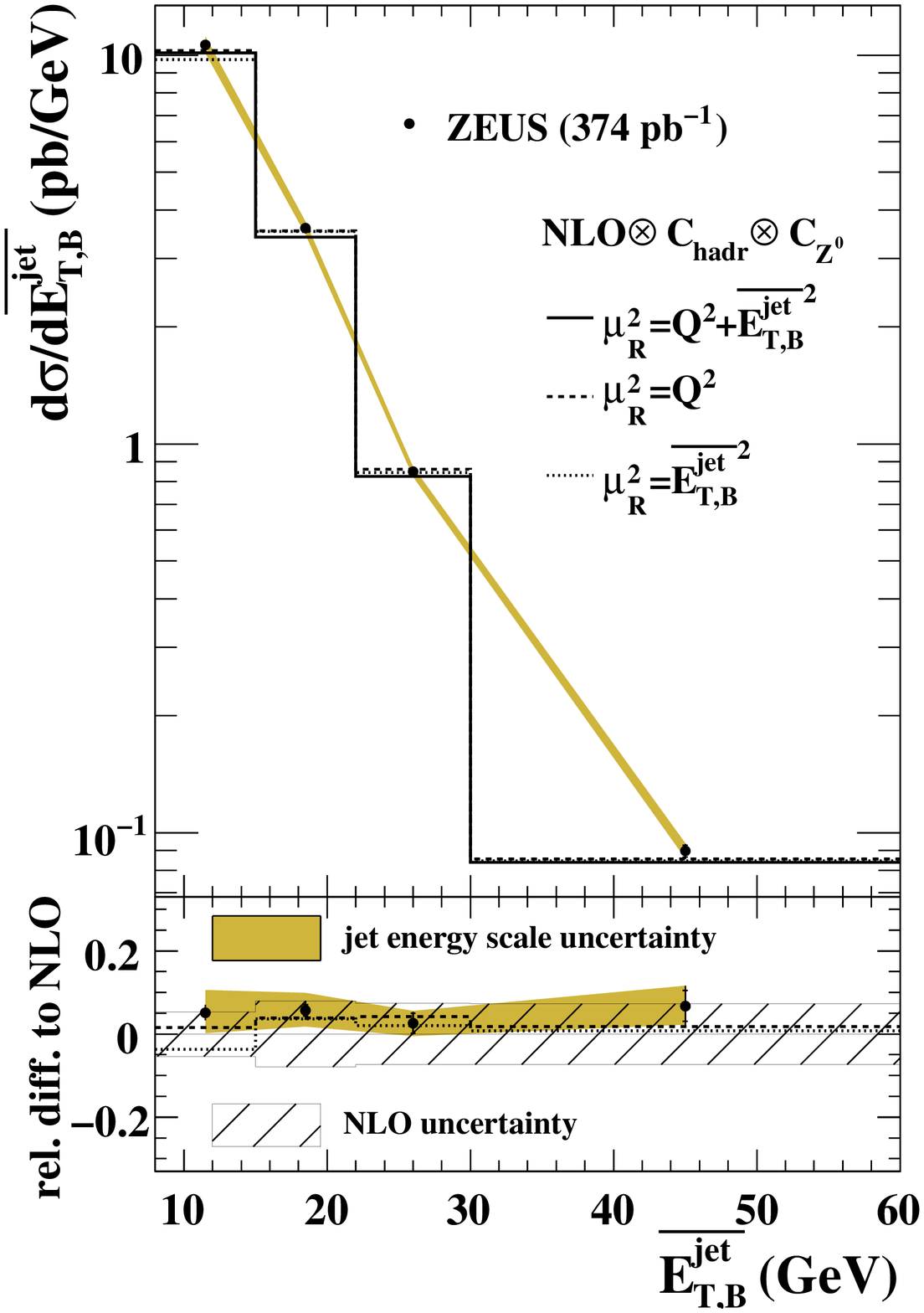,width=7.5cm}}
\put (8.5,0.){\epsfig{figure=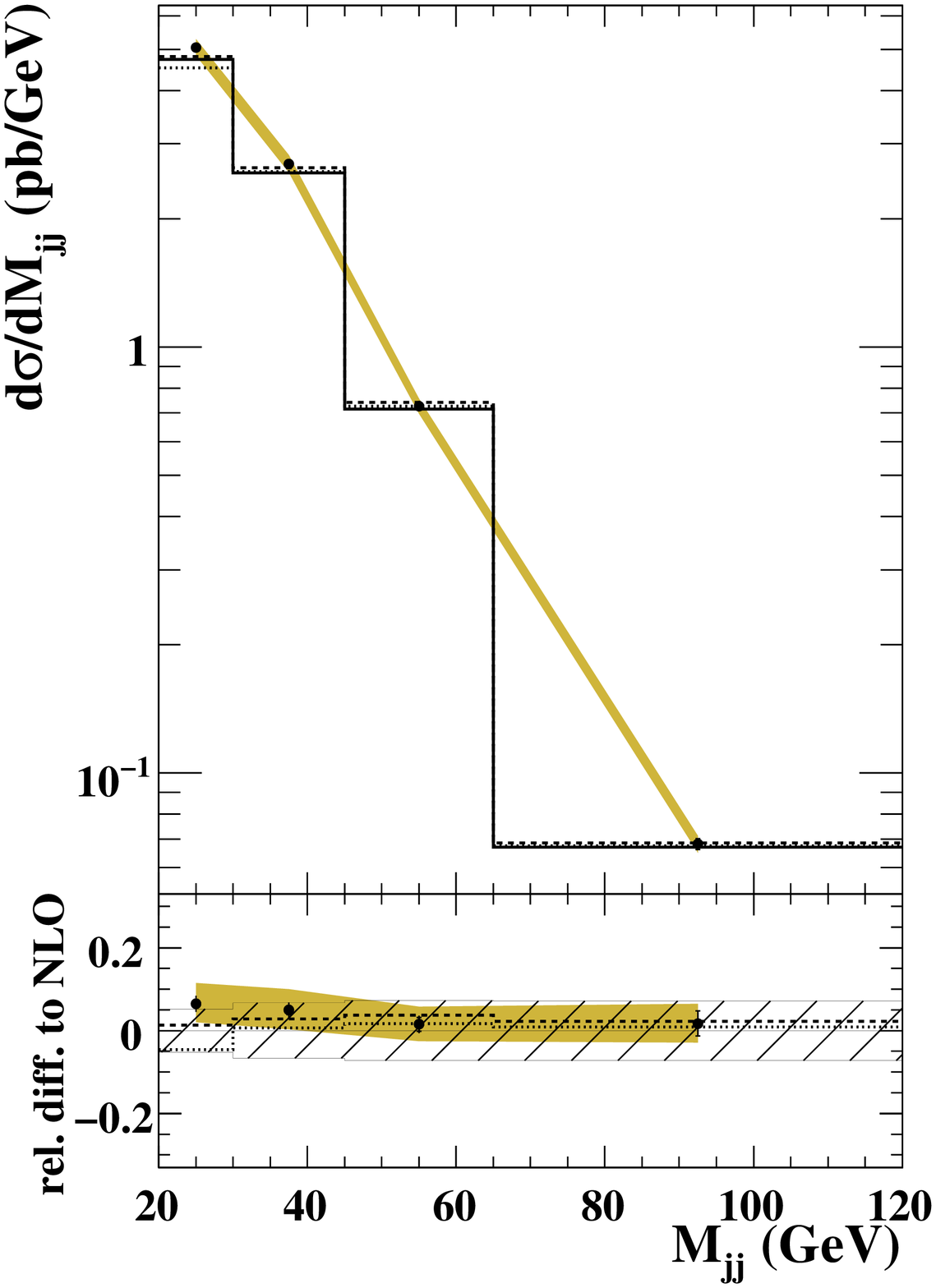,width=7.5cm}}
\put (7.25,10.5){{\bf {\Huge ZEUS}}}
\put (2.8,3.9){{\bf (a)}}
\put (10.3,3.9){{\bf (b)}}
\end{picture}
\vspace{1.cm}
\caption
{\it 
The measured differential cross-sections (a) $\mathit{d\sigma/d\overline{\etjb}}$ and 
(b) $\mathit{d\sigma/d\mj}$ for inclusive dijet production.
Other details as in the caption to
Fig.~\ref{fig1}. }
\label{fig2}
\vfill
\end{figure}

\begin{figure}[p]
\vfill
\setlength{\unitlength}{1.0cm}
\begin{picture} (18.0,15.0)
\put (1.0,0.0){\epsfig{figure=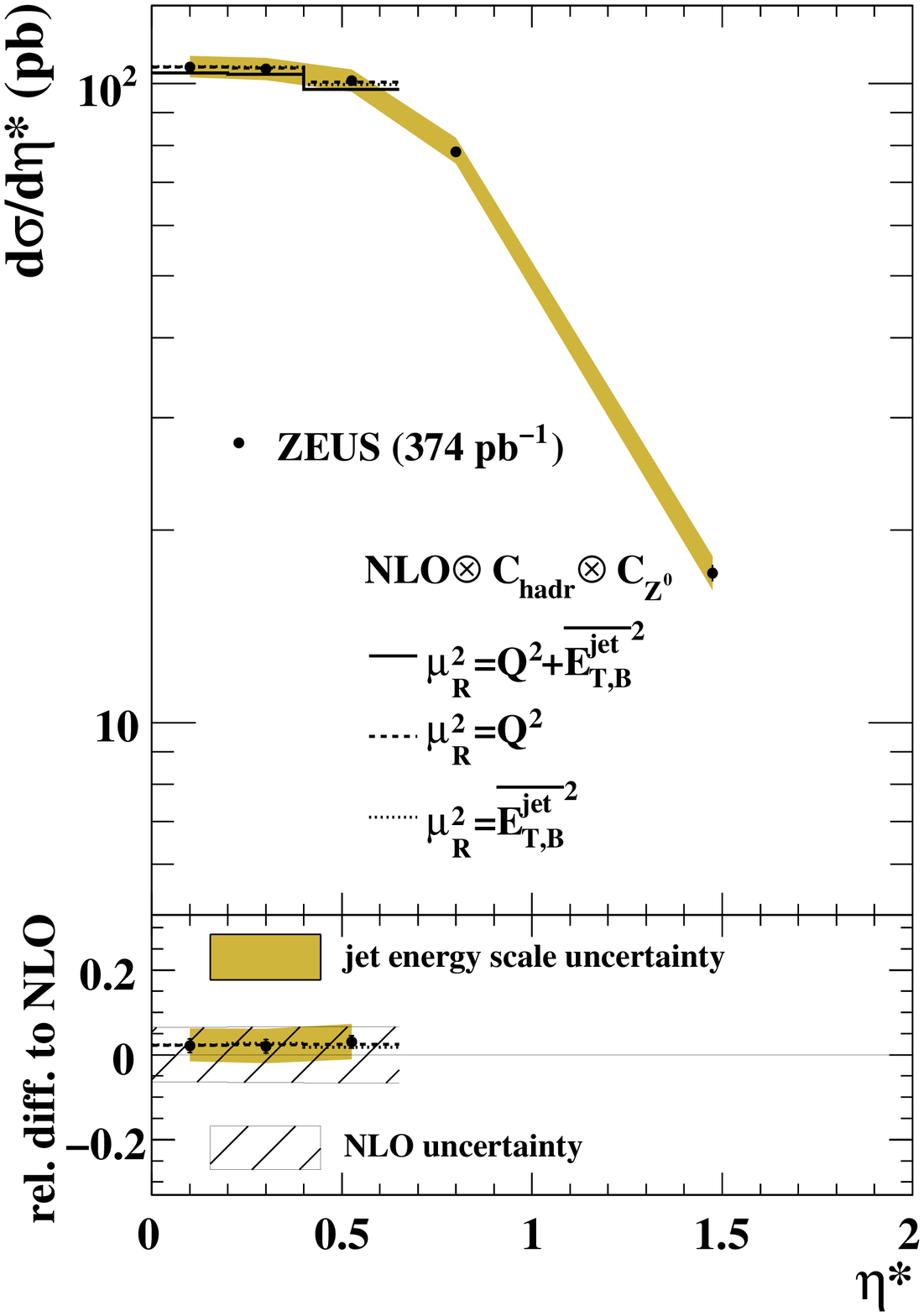,width=7.5cm}}
\put (8.5,0.){\epsfig{figure=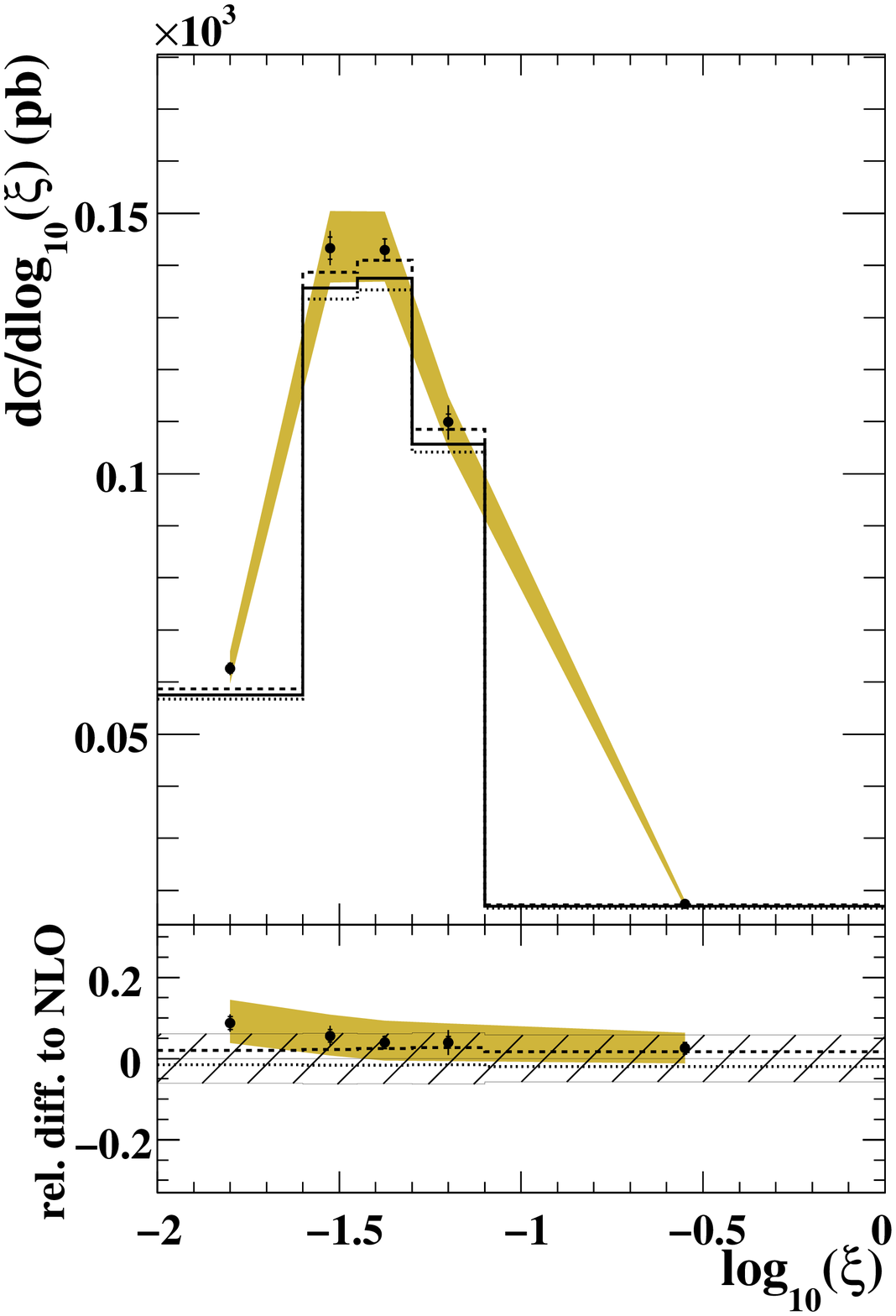,width=7.5cm}}
\put (7.25,10.5){{\bf {\Huge ZEUS}}}
\put (2.8,3.9){{\bf (a)}}
\put (10.3,3.9){{\bf (b)}}
\end{picture}
\vspace{1.cm}
\caption
{\it 
The measured differential cross-sections (a) $\mathit{d\sigma/d\eta^{*}}$ and (b) $\mathit{d\sigma/d\log_{10}\left(\xi\right)}$
for inclusive dijet production. In the last $\eta^{*}$ bins the NLO QCD predictions are not plotted for reasons explained in the text. 
Other details as in the caption to
Fig.~\ref{fig1}. }
\label{fig3}
\vfill
\end{figure}

The measurements of the single-differential inclusive dijet cross sections are presented in 
Figs~\ref{fig1} to~\ref{fig3} and Tables \ref{tab:q2} to~\ref{tab:xi} as functions of several kinematic and dijet variables. Single-differential cross sections are shown for $\q2$, $x_{\rm{Bj}}$, 
the mean transverse jet energy in the Breit frame of the two jets, $\overline{\etjb}$, the dijet invariant mass, $\mj$, 
the half-difference of the jet pseudorapidities in the Breit frame, $\eta^{*}$, and the 
logarithm of the variable $\xi$. The data are compared to NLO QCD calculations. 
The relative differences between the measured differential cross sections and the NLO QCD
calculations are also shown.

The single-differential dijet cross-sections $\sq2$ and $d\sigma/dx_{\rm{Bj}}$ are shown in Figs~\ref{fig1}a and \ref{fig1}b.
The cross-section $\sq2$ has total experimental systematic uncertainties of the order of 5$\%$ (7$\%$) at low (high) values of 
$\q2$. The total theoretical uncertainty is of the order of 7$\%$ (4$\%$) at low (high) $\q2$. 

For the cross-section $d\sigma/dx_{\rm{Bj}}$, most of the data points have 
experimental uncertainties of less than 5$\%$, and also the 
precision of the theory predictions is better than 5$\%$ over most of the $x_{\rm{Bj}}$
range. 

Figures~\ref{fig2}a and \ref{fig2}b show the single-differential dijet cross-sections 
$d\sigma/d\overline{\etjb}$ and $d\sigma/d\mj$. These measurements are 
particularly well suited for testing the matrix elements in the perturbative 
calculations. Mean transverse jet energies 
$\overline{\etjb}$ (dijet invariant masses $\mj$) of up to 60~GeV (120~GeV) are reached
with this measurement. At the largest values of $\overline{\etjb}$ ($\mj$), experimental
uncertainties of 8$\%$ (5$\%$) are achieved; for smaller values, the uncertainties are even 
smaller. The theoretical uncertainties are approximately constant over the range
studied and are of the order of $\left(5-7\right)\%$. 

The differential dijet cross-section as a function of $\eta^{*}$ is 
shown in Fig.~\ref{fig3}a. The experimental uncertainties 
are always below 5$\%$, the total theoretical uncertainty is also typically around 5$\%$. The theoretical predictions for the last 
two $\eta^{*}$ bins were removed 
from the plot due to infrared sensitivity.

The cross-section $d\sigma/d\log_{10}\left(\xi\right)$ (Fig.~\ref{fig3}b) has similar uncertainties as the distributions described before and 
shows a maximum around $\log_{10}\left(\xi\right)=-1.5$. At
lower and higher values, the cross section reflects the 
suppression by the transverse energy requirements in the selection and the decreasing
quark and gluon densities, respectively.

All the measured differential cross sections
are well described by NLO QCD predictions.
\subsection{Double-differential dijet cross sections}
\label{sec-results-double}
\begin{figure}[p]
\vfill
\centerline{\epsfig{figure=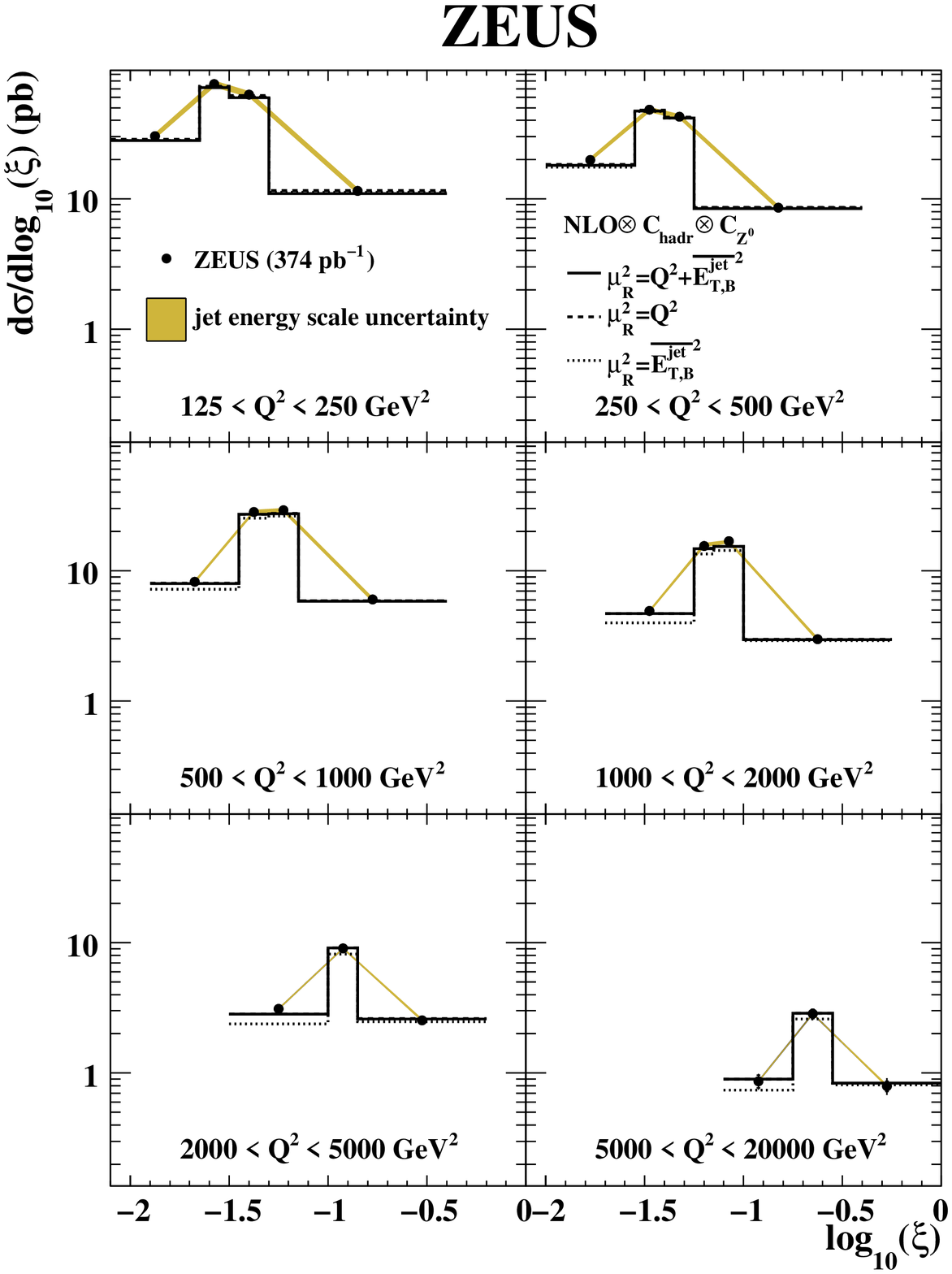,width=15.0cm}}
\caption
{\it
The measured differential cross-section $\mathit{d\sigma/d\log_{10}\left(\xi\right)}$ for
inclusive dijet production in different regions of $\mathit{\q2}$.
Other details as in the caption to Fig.~\ref{fig1}.
}
\label{fig4}
\vfill
\end{figure}

\begin{figure}[p]
\vfill
\centerline{\epsfig{figure=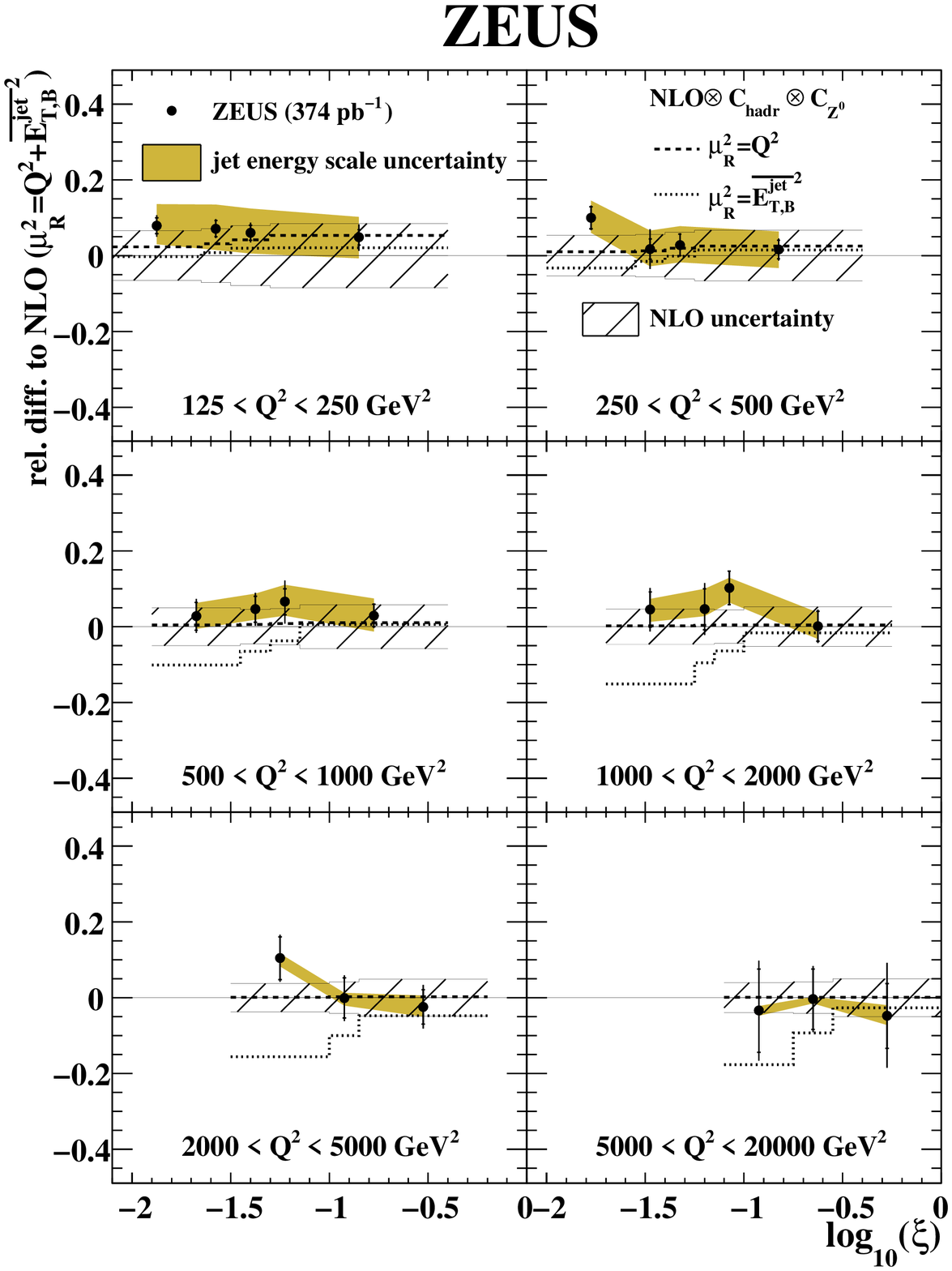,width=15.0cm}}
\caption
{\it
Relative differences between the measured differential cross-sections
$\mathit{d\sigma/d\log_{10}\left(\xi\right)}$ presented in Fig.~\ref{fig4} and the NLO QCD
calculations with $\mathit{\mu_R^2=\q2+\overline{\etjb}^2}$. Other details as in the caption to Fig.~\ref{fig1}.
}
\label{fig5}
\vfill
\end{figure}

\begin{figure}[p]
\vfill
\centerline{\epsfig{figure=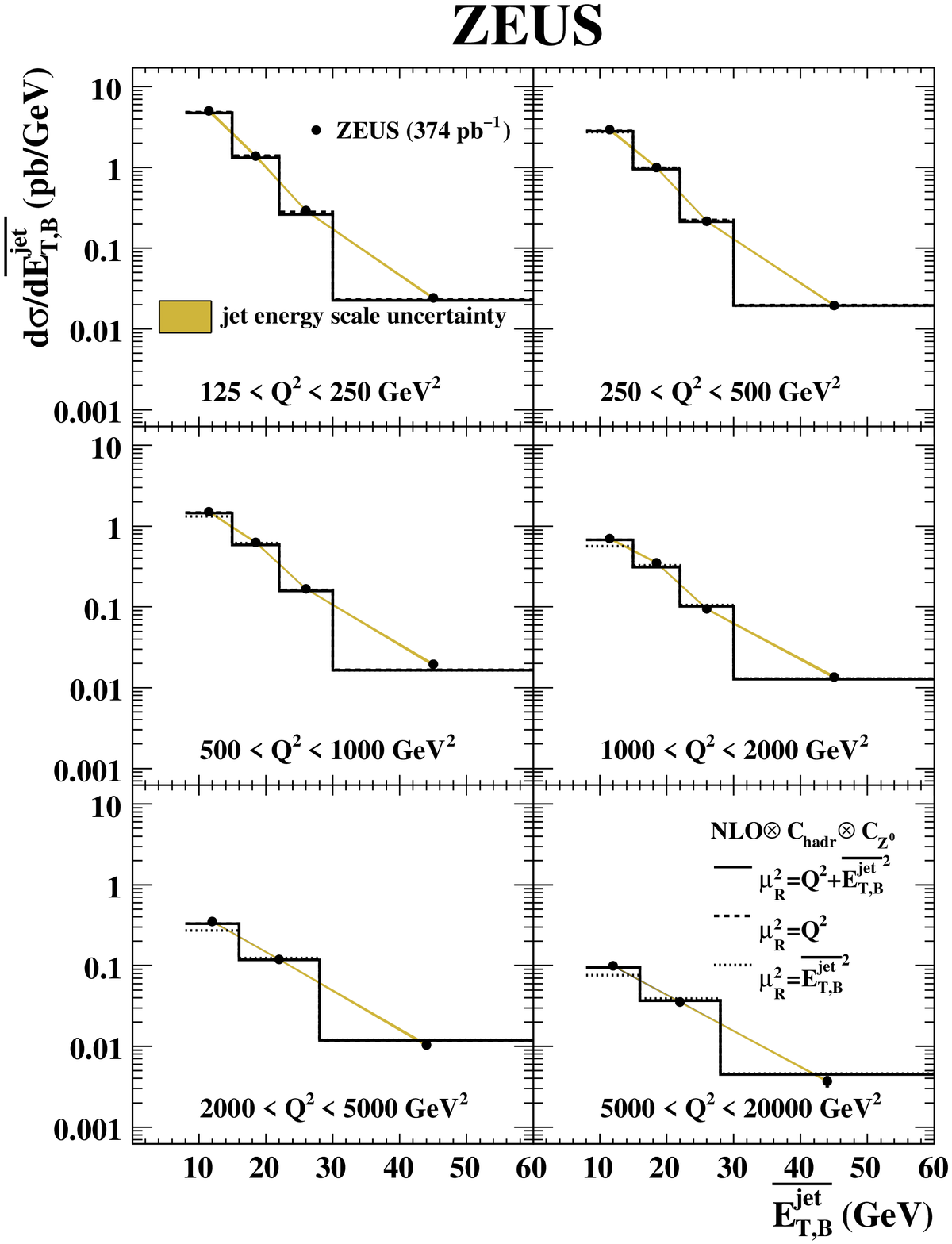,width=15.0cm}}
\caption
{\it
The measured differential cross-section $\mathit{d\sigma/d\overline{\etjb}}$ for
inclusive dijet production in different regions of $\mathit{\q2}$.
Other details as in the caption to Fig.~\ref{fig1}.
}
\label{fig6}
\vfill
\end{figure}

\begin{figure}[p]
\vfill
\centerline{\epsfig{figure=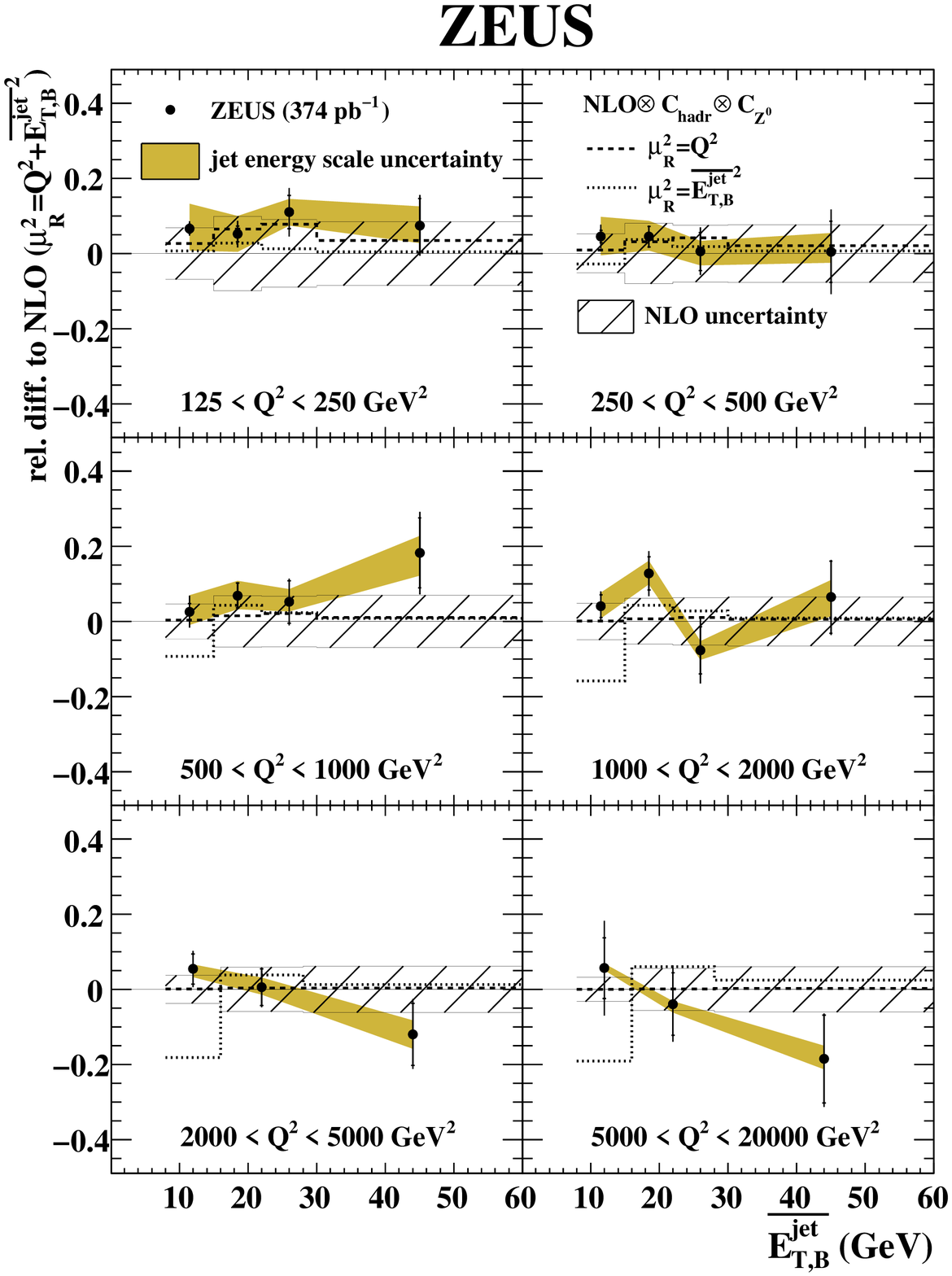,width=15.0cm}}
\caption
{\it
Relative differences between the measured differential cross-sections
$\mathit{d\sigma/d\overline{\etjb}}$ presented in Fig.~\ref{fig6} and the NLO QCD
calculations with $\mathit{\mu_R^2=\q2+\overline{\etjb}^2}$. Other details as in the caption to Fig.~\ref{fig1}.
}
\label{fig7}
\vfill
\end{figure}

Figures~\ref{fig4} to~\ref{fig7} show the measurements of double-differential
dijet cross sections as functions of $\overline{\etjb}$ and $\log_{10}\left(\xi\right)$ in different 
$\q2$ regions (see Tables \ref{tab:xi:dd} and \ref{tab:etmean:q2:dd}). These cross sections will provide valuable input for the extraction of the proton PDFs.

The $\log_{10}\left(\xi\right)$ distributions in different $\q2$ regions in Fig.~\ref{fig4} show the same 
behaviour as the integrated $\log_{10}\left(\xi\right)$ distribution in Fig.~\ref{fig3}b, with a distinct maximum
at values that increase with increasing $\q2$. The data are very precise -- even in the 
highest $\q2$ bin from 5\,000~to 20\,000~\g2 the experimental uncertainties are between 10
and 15$\%$ and originate equally from the statistical and the systematical uncertainty. 
At lower $\q2$ values, the experimental uncertainties become as small as $\left(2-3\right)\%$. 
Figure~\ref{fig5} shows the level of agreement between data and predictions: 
the theoretical uncertainties are typically between 5 and 10$\%$ and, within the combined 
uncertainties, the data are very well described by the theory.

The cross sections as functions of $\overline{\etjb}$ in 
different regions of $\q2$, shown in Fig.~\ref{fig6}, fall over 2 to 3 orders of magnitude 
in the range considered, with a smaller slope for higher $\q2$ values. The statistical
precision of the data is between 2$\%$ at the lowest $\overline{\etjb}$ and $\q2$ and slightly above 
10$\%$ at the highest values of these variables. The systematic uncertainties are mostly 
of the order of $3$ to $5\%$. The theoretical uncertainties (Fig.~\ref{fig7}) are approximately
constant in $\overline{\etjb}$; they are of the order of 5 to 10$\%$, with the
smaller values at higher $\q2$.
Data and theory are in good agreement over the whole measured range.

% ----------------------------------------------------------------------------
%        conclusion
% ----------------------------------------------------------------------------
\section{Summary and conclusions}
\label{sec-conclusion}
Measurements of single- and double-differential cross sections for dijet 
production at high-$\q2$ NC DIS were made using 
an integrated luminosity of 374~\pb1. The measurements have very small statistical and 
systematic uncertainties and the description of the data by the predictions of NLO QCD is 
very good, giving a powerful and stringent justification of the theory.
These data will provide useful precision information for the determination 
of the strong coupling constant and the extraction of the proton PDFs. 

\vspace{0.5cm}
\noindent {\Large\bf Acknowledgements}
\vspace{0.3cm}

We thank the DESY Directorate for their strong support and encouragement.
The remarkable achievements of the HERA machine group were essential for
the successful completion of this work and are greatly appreciated. We are
grateful for the support of the DESY computing and network services. The
design, construction and installation of the ZEUS detector have been made
possible owing to the ingenuity and effort of many people from DESY and home
institutes who are not listed as authors. We would like to thank Z. Nagy for 
useful discussions.

\vfill\eject

%
%------------------------------------------------------------------------------
%       Bibliography
%------------------------------------------------------------------------------
\raggedright
\providecommand{\etal}{et al.\xspace}
\providecommand{\coll}{Collaboration}
\catcode`\@=11
\def\@bibitem#1{%
\ifmc@bstsupport
  \mc@iftail{#1}%
    {;\newline\ignorespaces}%
    {\ifmc@first\else.\fi\orig@bibitem{#1}}
  \mc@firstfalse
\else
  \mc@iftail{#1}%
    {\ignorespaces}%
    {\orig@bibitem{#1}}%
\fi}%
\catcode`\@=12

\vfill\eject

%
%------------------------------------------------------------------------------
%       Tables
%------------------------------------------------------------------------------
%-------------------------------------
% put here all tables ...
%-------------------------------------
\begin{table}[ht]
\begin{center}
\renewcommand{\arraystretch}{1.25}
\begin{tabular}{||rll|llll||c||c||}
\hline
\multicolumn{3}{||l|}{$Q^{2}$ bin} & $d\sigma/dQ^{2}$ & & & & & \\
\multicolumn{3}{||l|}{$\left(\rm{GeV}^2\right)$} & $\left(\rm{pb/GeV^{2}}\right)$ & $\delta_{\rm{stat}}$ & $\delta_{\rm{syst}}$ & $\delta_{\rm{ES}}$ & $C_{\rm{QED}}$ & $C_{\rm{hadr}}\cdot C_{\rm{Z^{0}}}$\\
\hline \hline
$125$&$\ldots$&$250$ & $0.3843$ & $\pm0.0036$ & $^{+0.0039}_{-0.0040}$ & $^{+0.0215}_{-0.0195}$ &  $0.97$ & $0.95$ \\
$250$&$\ldots$&$500$ & $0.1193$ & $\pm0.0015$ & $^{+0.0019}_{-0.0018}$ & $^{+0.0055}_{-0.0052}$ &  $0.95$ & $0.96$ \\
$500$&$\ldots$&$1000$ & $0.03372$ & $\pm0.00053$ & $^{+0.00065}_{-0.00065}$ & $^{+0.00135}_{-0.00115}$ &  $0.94$ & $0.96$ \\
$1000$&$\ldots$&$2000$ & $0.00855$ & $\pm0.00018$ & $^{+0.00010}_{-0.00010}$ & $^{+0.00029}_{-0.00026}$ &  $0.93$ & $0.98$ \\
$2000$&$\ldots$&$5000$ & $0.001523$ & $\pm0.000043$ & $^{+0.000033}_{-0.000033}$ & $^{+0.000030}_{-0.000034}$ &  $0.93$ & $1.03$ \\
$5000$&$\ldots$&$20000$ & $0.0000875$ & $\pm0.0000046$ & $^{+0.0000058}_{-0.0000057}$ & $^{+0.0000014}_{-0.0000015}$ &  $0.92$ & $1.09$ \\
\hline
\end{tabular}
\caption{{\it The measured differential cross-sections d$\mathit{\sigma}$/d$\mathit{\q2}$ for inclusive dijet production. 
The statistical, uncorrelated systematic and jet-energy-scale (ES) uncertainties are shown separately. The multiplicative corrections, $\mathit{C_{\rm{QED}}}$, 
which have been applied to the data  
and the corrections for hadronisation and $\mathit{Z^{0}}$ effects to be applied to the 
parton-level NLO QCD calculations, $\mathit{C_{\rm{hadr}}\cdot C_{\rm{Z^{0}}}}$, are shown in the last two columns.}}
\label{tab:q2}
\end{center}
\end{table}

\begin{table}[ht]
\begin{center}
\renewcommand{\arraystretch}{1.25}
\begin{tabular}{||rll|llll||c||c||}
\hline
\multicolumn{3}{||l|}{$x_{\rm{Bj}}$ bin} & $d\sigma/dx_{\rm{Bj}}$ & & & & & \\
\multicolumn{3}{||l|}{} & $\left(\rm{pb}\right)$ & $\delta_{\rm{stat}}$ & $\delta_{\rm{syst}}$ & $\delta_{\rm{ES}}$ & $C_{\rm{QED}}$ & $C_{\rm{hadr}}\cdot C_{\rm{Z^{0}}}$\\
\hline \hline
$0.0001$&$\ldots$&$0.01$ & $6580$ & $\pm54$ & $^{+44}_{-45}$ & $^{+351}_{-317}$ &  $0.96$ & $0.95$ \\
$0.01$&$\ldots$&$0.02$ & $2229$ & $\pm31$ & $^{+44}_{-44}$ & $^{+98}_{-94}$ &  $0.94$ & $0.95$ \\
$0.02$&$\ldots$&$0.035$ & $711$ & $\pm14$ & $^{+19}_{-20}$ & $^{+27}_{-23}$ &  $0.94$ & $0.96$ \\
$0.035$&$\ldots$&$0.07$ & $193.8$ & $\pm4.6$ & $^{+2.8}_{-2.5}$ & $^{+6.1}_{-5.7}$ &  $0.93$ & $0.99$ \\
$0.07$&$\ldots$&$0.1$ & $64.4$ & $\pm2.8$ & $^{+3.8}_{-3.8}$ & $^{+1.0}_{-1.4}$ &  $0.92$ & $1.03$ \\
\hline
\end{tabular}
\caption{{\it Inclusive dijet cross-sections $\mathit{d\sigma/dx_{\rm{Bj}}}$. Other details as in the caption to Table~\ref{tab:q2}.}}
\label{tab:xbj}
\end{center}
\end{table}

\begin{table}[ht]
\begin{center}
\renewcommand{\arraystretch}{1.25}
\begin{tabular}{||rll|llll||c||c||}
\hline
\multicolumn{3}{||l|}{$\overline{\etjb}$ bin} & $d\sigma/d\overline{\etjb}$ & & & & & \\
\multicolumn{3}{||l|}{$\left(\rm{GeV}\right)$} & $\left(\rm{pb/GeV}\right)$ & $\delta_{\rm{stat}}$ & $\delta_{\rm{syst}}$ & $\delta_{\rm{ES}}$ & $C_{\rm{QED}}$ & $C_{\rm{hadr}}\cdot C_{\rm{Z^{0}}}$\\
\hline \hline
$8$&$\ldots$&$15$ & $10.650$ & $\pm0.083$ & $^{+0.174}_{-0.174}$ & $^{+0.549}_{-0.495}$ &  $0.95$ & $0.95$ \\
$15$&$\ldots$&$22$ & $3.595$ & $\pm0.046$ & $^{+0.060}_{-0.062}$ & $^{+0.142}_{-0.134}$ &  $0.95$ & $0.98$ \\
$22$&$\ldots$&$30$ & $0.848$ & $\pm0.020$ & $^{+0.011}_{-0.010}$ & $^{+0.025}_{-0.026}$ &  $0.95$ & $0.99$ \\
$30$&$\ldots$&$60$ & $0.0896$ & $\pm0.0031$ & $^{+0.0027}_{-0.0027}$ & $^{+0.0041}_{-0.0038}$ &  $0.95$ & $0.99$ \\
\hline
\end{tabular}
\caption{{\it Inclusive dijet cross-sections $\mathit{d\sigma/d\overline{\etjb}}$. Other details as in the caption to Table~\ref{tab:q2}.}}
\label{tab:etmean}
\end{center}
\end{table}

\begin{table}[ht]
\begin{center}
\renewcommand{\arraystretch}{1.25}
\begin{tabular}{||rll|llll||c||c||}
\hline
\multicolumn{3}{||l|}{$\mj$ bin} & $d\sigma/d\mj$ & & & & & \\
\multicolumn{3}{||l|}{$\left(\rm{GeV}\right)$} & $\left(\rm{pb/GeV}\right)$ & $\delta_{\rm{stat}}$ & $\delta_{\rm{syst}}$ & $\delta_{\rm{ES}}$ & $C_{\rm{QED}}$ & $C_{\rm{hadr}}\cdot C_{\rm{Z^{0}}}$\\
\hline \hline
$20$&$\ldots$&$30$ & $5.048$ & $\pm0.049$ & $^{+0.079}_{-0.079}$ & $^{+0.236}_{-0.212}$ &  $0.95$ & $0.95$ \\
$30$&$\ldots$&$45$ & $2.693$ & $\pm0.028$ & $^{+0.038}_{-0.038}$ & $^{+0.130}_{-0.121}$ &  $0.95$ & $0.97$ \\
$45$&$\ldots$&$65$ & $0.726$ & $\pm0.012$ & $^{+0.009}_{-0.010}$ & $^{+0.031}_{-0.029}$ &  $0.95$ & $0.98$ \\
$65$&$\ldots$&$120$ & $0.0681$ & $\pm0.0020$ & $^{+0.0005}_{-0.0005}$ & $^{+0.0032}_{-0.0031}$ &  $0.95$ & $0.97$ \\
\hline
\end{tabular}
\caption{{\it Inclusive dijet cross-sections $\mathit{d\sigma/d\mj}$. Other details as in the caption to Table~\ref{tab:q2}.}}
\label{tab:m}
\end{center}
\end{table}

\begin{table}[ht]
\begin{center}
\renewcommand{\arraystretch}{1.25}
\begin{tabular}{||rll|llll||c||c||}
\hline
\multicolumn{3}{||l|}{$\eta^{*}$ bin} & $d\sigma/d\eta^{*}$ & & & & & \\
\multicolumn{3}{||l|}{} & $\left(\rm{pb}\right)$ & $\delta_{\rm{stat}}$ & $\delta_{\rm{syst}}$ & $\delta_{\rm{ES}}$ & $C_{\rm{QED}}$ & $C_{\rm{hadr}}\cdot C_{\rm{Z^{0}}}$\\
\hline \hline
$0$&$\ldots$&$0.2$ & $106.1$ & $\pm1.6$ & $^{+0.9}_{-0.8}$ & $^{+4.3}_{-3.8}$ &  $0.95$ & $0.96$ \\
$0.2$&$\ldots$&$0.4$ & $105.4$ & $\pm1.6$ & $^{+0.9}_{-0.9}$ & $^{+4.3}_{-4.1}$ &  $0.95$ & $0.96$ \\
$0.4$&$\ldots$&$0.65$ & $101.0$ & $\pm1.4$ & $^{+0.6}_{-0.7}$ & $^{+4.1}_{-4.0}$ &  $0.96$ & $0.97$ \\
$0.65$&$\ldots$&$0.95$ & $78.2$ & $\pm1.1$ & $^{+0.3}_{-0.4}$ & $^{+4.0}_{-3.3}$ &  $0.95$ & $0.98$ \\
$0.95$&$\ldots$&$2$ & $17.14$ & $\pm0.27$ & $^{+0.42}_{-0.42}$ & $^{+1.06}_{-1.02}$ &  $0.95$ & $0.93$ \\
\hline
\end{tabular}
\caption{{\it Inclusive dijet cross-sections $\mathit{d\sigma/d\eta^{*}}$. Other details as in the caption to Table~\ref{tab:q2}.}}
\label{tab:eta}
\end{center}
\end{table}

\begin{table}[ht]
\begin{center}
\renewcommand{\arraystretch}{1.25}
\begin{tabular}{||rll|llll||c||c||}
\hline
\multicolumn{3}{||l|}{$\log_{10}\left(\xi\right)$ bin} & $d\sigma/d\log_{10}\left(\xi\right)$ & & & & & \\
\multicolumn{3}{||l|}{} & $\left(\rm{pb}\right)$ & $\delta_{\rm{stat}}$ & $\delta_{\rm{syst}}$ & $\delta_{\rm{ES}}$ & $C_{\rm{QED}}$ & $C_{\rm{hadr}}\cdot C_{\rm{Z^{0}}}$\\
\hline \hline
$-2$&$\ldots$&$-1.6$ & $62.63$ & $\pm0.91$ & $^{+0.81}_{-0.86}$ & $^{+3.27}_{-2.83}$ &  $0.97$ & $0.95$ \\
$-1.6$&$\ldots$&$-1.45$ & $143.3$ & $\pm2.1$ & $^{+2.5}_{-2.5}$ & $^{+7.1}_{-6.6}$ &  $0.96$ & $0.95$ \\
$-1.45$&$\ldots$&$-1.3$ & $143.0$ & $\pm2.1$ & $^{+1.0}_{-0.8}$ & $^{+7.3}_{-6.1}$ &  $0.95$ & $0.96$ \\
$-1.3$&$\ldots$&$-1.1$ & $109.9$ & $\pm1.5$ & $^{+2.9}_{-3.0}$ & $^{+4.9}_{-4.9}$ &  $0.95$ & $0.96$ \\
$-1.1$&$\ldots$&$0$ & $17.40$ & $\pm0.24$ & $^{+0.11}_{-0.13}$ & $^{+0.65}_{-0.63}$ &  $0.94$ & $0.98$ \\
\hline
\end{tabular}
\caption{{\it Inclusive dijet cross-sections $\mathit{d\sigma/d\log_{10}\left(\xi\right)}$. Other details as in the caption to Table~\ref{tab:q2}.}}
\label{tab:xi}
\end{center}
\end{table}

\begin{table}
\begin{center}
\renewcommand{\arraystretch}{1.25}
\begin{tabular}{||rll|llll||c||c||}
\hline
\multicolumn{3}{||l|}{$\log_{10}(\xi)$ bin} & $d\sigma/d\log_{10}\left(\xi\right)$ & & & & & \\
 & & & $\left(\rm{pb}\right)$ & $\delta_{\rm{stat}}$ & $\delta_{\rm{syst}}$ & $\delta_{\rm{ES}}$ & $C_{\rm{QED}}$ & $C_{\rm{hadr}}\cdot C_{\rm{Z^{0}}}$\\
\hline \hline
\multicolumn{9}{||c||}{$125 < Q^{2} < 250~\rm{GeV}^{2}$}\\
\hline
$-2.1$&$\ldots$&$-1.65$ & $30.34$ & $\pm0.60$ & $^{+0.51}_{-0.55}$ & $^{+1.58}_{-1.38}$ &  $0.97$ & $0.95$ \\
$-1.65$&$\ldots$&$-1.5$ & $76.2$ & $\pm1.5$ & $^{+1.0}_{-0.9}$ & $^{+4.5}_{-4.0}$ &  $0.96$ & $0.94$ \\
$-1.5$&$\ldots$&$-1.3$ & $63.2$ & $\pm1.2$ & $^{+1.1}_{-1.0}$ & $^{+3.8}_{-3.3}$ &  $0.97$ & $0.97$ \\
$-1.3$&$\ldots$&$-0.4$ & $11.53$ & $\pm0.22$ & $^{+0.31}_{-0.33}$ & $^{+0.60}_{-0.61}$ &  $0.97$ & $0.94$ \\
\hline
\multicolumn{9}{||c||}{$250 < Q^{2} < 500~\rm{GeV}^{2}$}\\
\hline
$-2$&$\ldots$&$-1.55$ & $19.93$ & $\pm0.52$ & $^{+0.29}_{-0.27}$ & $^{+0.82}_{-0.71}$ &  $0.96$ & $0.95$ \\
$-1.55$&$\ldots$&$-1.4$ & $48.2$ & $\pm1.3$ & $^{+2.1}_{-2.1}$ & $^{+2.3}_{-2.2}$ &  $0.94$ & $0.96$ \\
$-1.4$&$\ldots$&$-1.25$ & $42.8$ & $\pm1.2$ & $^{+0.5}_{-0.4}$ & $^{+2.1}_{-1.9}$ &  $0.95$ & $0.96$ \\
$-1.25$&$\ldots$&$-0.4$ & $8.56$ & $\pm0.21$ & $^{+0.13}_{-0.14}$ & $^{+0.40}_{-0.41}$ &  $0.95$ & $0.95$ \\
\hline
\multicolumn{9}{||c||}{$500 < Q^{2} < 1000~\rm{GeV}^{2}$}\\
\hline
$-1.9$&$\ldots$&$-1.45$ & $8.21$ & $\pm0.29$ & $^{+0.22}_{-0.19}$ & $^{+0.29}_{-0.27}$ &  $0.94$ & $0.94$ \\
$-1.45$&$\ldots$&$-1.3$ & $28.32$ & $\pm0.93$ & $^{+0.74}_{-0.72}$ & $^{+1.10}_{-0.76}$ &  $0.94$ & $0.96$ \\
$-1.3$&$\ldots$&$-1.15$ & $29.11$ & $\pm0.93$ & $^{+1.23}_{-1.27}$ & $^{+1.22}_{-1.02}$ &  $0.94$ & $0.97$ \\
$-1.15$&$\ldots$&$-0.4$ & $6.00$ & $\pm0.17$ & $^{+0.07}_{-0.08}$ & $^{+0.26}_{-0.25}$ &  $0.95$ & $0.97$ \\
\hline
\multicolumn{9}{||c||}{$1000 < Q^{2} < 2000~\rm{GeV}^{2}$}\\
\hline
$-1.7$&$\ldots$&$-1.25$ & $4.91$ & $\pm0.22$ & $^{+0.16}_{-0.16}$ & $^{+0.13}_{-0.15}$ &  $0.94$ & $0.95$ \\
$-1.25$&$\ldots$&$-1.15$ & $15.52$ & $\pm0.80$ & $^{+0.64}_{-0.63}$ & $^{+0.80}_{-0.27}$ &  $0.93$ & $0.98$ \\
$-1.15$&$\ldots$&$-1$ & $16.87$ & $\pm0.68$ & $^{+0.10}_{-0.15}$ & $^{+0.41}_{-0.58}$ &  $0.94$ & $0.98$ \\
$-1$&$\ldots$&$-0.25$ & $2.97$ & $\pm0.12$ & $^{+0.06}_{-0.06}$ & $^{+0.11}_{-0.10}$ &  $0.93$ & $1.00$ \\
\hline
\multicolumn{9}{||c||}{$2000 < Q^{2} < 5000~\rm{GeV}^{2}$}\\
\hline
$-1.5$&$\ldots$&$-1$ & $3.11$ & $\pm0.16$ & $^{+0.07}_{-0.08}$ & $^{+0.04}_{-0.06}$ &  $0.93$ & $1.03$ \\
$-1$&$\ldots$&$-0.85$ & $9.07$ & $\pm0.48$ & $^{+0.26}_{-0.28}$ & $^{+0.12}_{-0.18}$ &  $0.92$ & $1.03$ \\
$-0.85$&$\ldots$&$-0.2$ & $2.54$ & $\pm0.12$ & $^{+0.09}_{-0.09}$ & $^{+0.08}_{-0.07}$ &  $0.93$ & $1.04$ \\
\hline
\multicolumn{9}{||c||}{$5000 < Q^{2} < 20000~\rm{GeV}^{2}$}\\
\hline
$-1.1$&$\ldots$&$-0.75$ & $0.865$ & $\pm0.099$ & $^{+0.065}_{-0.065}$ & $^{+0.011}_{-0.013}$ &  $0.94$ & $1.10$ \\
$-0.75$&$\ldots$&$-0.55$ & $2.85$ & $\pm0.23$ & $^{+0.10}_{-0.10}$ & $^{+0.02}_{-0.03}$ &  $0.91$ & $1.08$ \\
$-0.55$&$\ldots$&$0$ & $0.794$ & $\pm0.071$ & $^{+0.092}_{-0.090}$ & $^{+0.023}_{-0.020}$ &  $0.92$ & $1.10$ \\
\hline
\end{tabular}
\caption{{\it Inclusive dijet cross-sections $\mathit{d\sigma/d\log_{10}\left(\xi\right)}$
in different regions of $\mathit{Q^{2}}$. Other details as in the caption to Table~\ref{tab:q2}.}}
\label{tab:xi:dd}
\end{center}
\end{table}

\begin{table}
\begin{center}
\renewcommand{\arraystretch}{1.25}
\begin{tabular}{||rll|llll||c||c||}
\hline
\multicolumn{3}{||l|}{$\overline{\etjb}$ bin} & $d\sigma/\overline{\etjb}$ & & & & & \\
\multicolumn{3}{||l|}{$\left(\rm{GeV}\right)$} & $\left(\rm{pb/GeV}\right)$ & $\delta_{\rm{stat}}$ & $\delta_{\rm{syst}}$ & $\delta_{\rm{ES}}$ & $C_{\rm{QED}}$ & $C_{\rm{hadr}}\cdot C_{\rm{Z^{0}}}$\\
\hline \hline
\multicolumn{9}{||c||}{$125 < Q^{2} < 250~\rm{GeV}^{2}$}\\
\hline
$8$&$\ldots$&$15$ & $5.050$ & $\pm0.057$ & $^{+0.071}_{-0.070}$ & $^{+0.311}_{-0.274}$ &  $0.97$ & $0.95$ \\
$15$&$\ldots$&$22$ & $1.385$ & $\pm0.028$ & $^{+0.037}_{-0.038}$ & $^{+0.063}_{-0.062}$ &  $0.97$ & $0.96$ \\
$22$&$\ldots$&$30$ & $0.292$ & $\pm0.012$ & $^{+0.012}_{-0.013}$ & $^{+0.009}_{-0.010}$ &  $0.97$ & $0.96$ \\
$30$&$\ldots$&$60$ & $0.0241$ & $\pm0.0016$ & $^{+0.0009}_{-0.0008}$ & $^{+0.0011}_{-0.0010}$ &  $0.97$ & $0.95$ \\
\hline
\multicolumn{9}{||c||}{$250 < Q^{2} < 500~\rm{GeV}^{2}$}\\
\hline
$8$&$\ldots$&$15$ & $2.937$ & $\pm0.046$ & $^{+0.076}_{-0.076}$ & $^{+0.146}_{-0.141}$ &  $0.95$ & $0.95$ \\
$15$&$\ldots$&$22$ & $0.998$ & $\pm0.026$ & $^{+0.011}_{-0.011}$ & $^{+0.040}_{-0.035}$ &  $0.95$ & $0.98$ \\
$22$&$\ldots$&$30$ & $0.215$ & $\pm0.011$ & $^{+0.008}_{-0.008}$ & $^{+0.006}_{-0.008}$ &  $0.96$ & $0.97$ \\
$30$&$\ldots$&$60$ & $0.0195$ & $\pm0.0016$ & $^{+0.0015}_{-0.0015}$ & $^{+0.0010}_{-0.0006}$ &  $0.93$ & $0.95$ \\
\hline
\multicolumn{9}{||c||}{$500 < Q^{2} < 1000~\rm{GeV}^{2}$}\\
\hline
$8$&$\ldots$&$15$ & $1.502$ & $\pm0.031$ & $^{+0.055}_{-0.054}$ & $^{+0.064}_{-0.052}$ &  $0.94$ & $0.95$ \\
$15$&$\ldots$&$22$ & $0.629$ & $\pm0.019$ & $^{+0.008}_{-0.009}$ & $^{+0.023}_{-0.021}$ &  $0.95$ & $0.99$ \\
$22$&$\ldots$&$30$ & $0.1665$ & $\pm0.0089$ & $^{+0.0041}_{-0.0041}$ & $^{+0.0054}_{-0.0040}$ &  $0.96$ & $0.98$ \\
$30$&$\ldots$&$60$ & $0.0194$ & $\pm0.0015$ & $^{+0.0010}_{-0.0010}$ & $^{+0.0007}_{-0.0010}$ &  $0.95$ & $0.99$ \\
\hline
\multicolumn{9}{||c||}{$1000 < Q^{2} < 2000~\rm{GeV}^{2}$}\\
\hline
$8$&$\ldots$&$15$ & $0.701$ & $\pm0.020$ & $^{+0.017}_{-0.017}$ & $^{+0.025}_{-0.022}$ &  $0.93$ & $0.95$ \\
$15$&$\ldots$&$22$ & $0.352$ & $\pm0.014$ & $^{+0.012}_{-0.013}$ & $^{+0.011}_{-0.009}$ &  $0.94$ & $1.01$ \\
$22$&$\ldots$&$30$ & $0.0943$ & $\pm0.0064$ & $^{+0.0063}_{-0.0063}$ & $^{+0.0025}_{-0.0026}$ &  $0.94$ & $1.02$ \\
$30$&$\ldots$&$60$ & $0.0136$ & $\pm0.0012$ & $^{+0.0003}_{-0.0004}$ & $^{+0.0006}_{-0.0007}$ &  $0.94$ & $1.04$ \\
\hline
\multicolumn{9}{||c||}{$2000 < Q^{2} < 5000~\rm{GeV}^{2}$}\\
\hline
$8$&$\ldots$&$16$ & $0.350$ & $\pm0.013$ & $^{+0.009}_{-0.009}$ & $^{+0.004}_{-0.007}$ &  $0.92$ & $1.00$ \\
$16$&$\ldots$&$28$ & $0.1191$ & $\pm0.0058$ & $^{+0.0023}_{-0.0022}$ & $^{+0.0030}_{-0.0023}$ &  $0.93$ & $1.07$ \\
$28$&$\ldots$&$60$ & $0.01040$ & $\pm0.00097$ & $^{+0.00053}_{-0.00049}$ & $^{+0.00044}_{-0.00046}$ &  $0.94$ & $1.08$ \\
\hline
\multicolumn{9}{||c||}{$5000 < Q^{2} < 20000~\rm{GeV}^{2}$}\\
\hline
$8$&$\ldots$&$16$ & $0.0995$ & $\pm0.0076$ & $^{+0.0092}_{-0.0092}$ & $^{+0.0012}_{-0.0005}$ &  $0.93$ & $1.05$ \\
$16$&$\ldots$&$28$ & $0.0354$ & $\pm0.0031$ & $^{+0.0023}_{-0.0021}$ & $^{+0.0003}_{-0.0008}$ &  $0.89$ & $1.14$ \\
$28$&$\ldots$&$60$ & $0.00368$ & $\pm0.00053$ & $^{+0.00015}_{-0.00023}$ & $^{+0.00016}_{-0.00012}$ &  $0.95$ & $1.20$ \\
\hline
\end{tabular}
\caption{{\it Inclusive dijet cross-sections $\mathit{d\sigma/d\overline{\etjb}}$
in different regions of $\mathit{Q^{2}}$. Other details as in the caption to Table~\ref{tab:q2}.}}
\label{tab:etmean:q2:dd}
\end{center}
\end{table}

%
%------------------------------------------------------------------------------
%       Figures
%------------------------------------------------------------------------------
%-------------------------------------------------------------------------------
%       Results: the figures
%-------------------------------------------------------------------------------
\newpage
\clearpage

\newpage
\clearpage

\newpage
\clearpage

\newpage
\clearpage

\newpage
\clearpage

\newpage
\clearpage
\newpage
\clearpage
\newpage
\clearpage

\newpage
\clearpage

\newpage
\clearpage

\newpage
\clearpage

\newpage
\clearpage

\newpage
\clearpage

\newpage
\clearpage

%
%       ... that's it
%
\end{document}